\documentclass{andromedaone}       

\journal{BSM}%Letters in High Energy Physics}
\vol{2021}
\jyear{Egypt}
\pages{Zewali City} 

\received{xx April 2021}
\published{xx June 2021}

\usepackage[italian, english]{babel}
\usepackage{ifpdf}
\usepackage{graphics, graphicx, subfigure}
\usepackage{amsmath, amssymb, amsfonts, nicefrac, slashed, cancel}
\usepackage{epsf}
\usepackage{rotating}
\usepackage{fancyhdr}
\usepackage{lineno}
\usepackage{pstricks}
\usepackage{color, xcolor}
\usepackage{multirow}
\usepackage{bm}
\usepackage{float}
\usepackage{array}
\usepackage{boldline}
\usepackage{mathrsfs}
\usepackage{stackengine}

\usepackage{hyperref}
\hypersetup{colorlinks=true,linkcolor=blue,citecolor=red,urlcolor=lightseagreen}

\usepackage{tikz}
\usepackage{tikz-feynman}
\tikzfeynmanset{compat=1.0.0}

\newcommand*\xbar[1]{%
	\hbox{%
		\vbox{%
			\hrule height 0.65pt % The actual bar
			\kern0.4ex%         % Distance between bar and symbol
			\hbox{%
				\kern-0.05em%      % Shortening on the left side
				\ensuremath{#1}%
				\kern0.0em%      % Shortening on the right side
			}%
		}%
	}%
}

\def\be{\begin{equation}}
\def\ee{\end{equation}}
\def\bea{\begin{eqnarray}}
\def\eea{\end{eqnarray}}

\DeclareFontFamily{OT1}{pzc}{}
\DeclareFontShape{OT1}{pzc}{m}{it}{<-> s * [1.2] pzcmi7t}{}
\DeclareMathAlphabet{\mathpzc}{OT1}{pzc}{m}{it}
\newcommand{\Q}{\bm{Q}}
\newcommand{\LL}{\bm{L}}
\newcommand{\cl}{\mathpzc{l}}
\newcommand{\uq}{\mathpzc{u}}
\newcommand{\dq}{\mathpzc{d}}

\begin{document}

\title{Distinguishing Different BSM Signatures at Present and Future Colliders}

\author{Saunak Dutta\auno{1}, Priyotosh Bandyopadhyay\auno{1} and Anirban Karan\auno{1}}
\address{$^1$ Indian Institute of Technology Hyderabad, Telangana 502285}

\begin{abstract}
We show how angular distributions can distinguish different scenarios beyond the standard model by characterising particles of different spins at the LHC. We illustrate the idea with scalar and vector leptoquarks along with the heavy fermions in Type-III seesaw as spin zero, spin one and spin half examples respectively. On the other hand, zeros of single photon tree level amplitude can separate different particles according their electromagnetic charges. This phenomenon can be used to distinguish leptoquarks of different gauge representations, even different excitations of same $SU(2)_L$ gauge group, within the same spin frame work. We explore electron-photon and electron-hadron colliders to discern such scenarios in the context of the leptoquark models by means of zeros in scattering amplitudes. We found that the discerning effect in these two colliders are complementary to each other and both of them are required for an exhaustive analysis of leptoquark models. The analyses are carried out for different leptoquark masses and centre of mass energies of the collisions which involve a PYTHIA based simulation.
\end{abstract}

\maketitle


%%%%%%%%%%%%%%%%%%%%%%%%%%%%%%%%%%%%%%%%%%%%%%%%%%%%%%%%%%%%%%%%%%%%%%%%%%%%%%%%%%%%%%%%
\section{Introduction}
The Standard Model of Particle Physics (SM), despite its accomplishment in unifying the fundamental constituents and interactions of nature, fails to address several observational discrepancies and suffers from theoretical inconsistencies in its own framework. Different theories Beyond the Standard Model (BSM) involving various additional particles have been proposed so far to address them. However, it is a challenging task to identify them in particle colliders, well distinguished from SM Background as well as other BSM theories. In this article, we have shown how different BSM models can be classified based on spins and gauge representations of the new particles they propose.

 As an illustration, we consider the leptoquarks which appear with spins-0 and 1, along with singlet, doublet and triplet $SU(2)_L$ representations. We show that the spin of any leptoquark can be distinguished in its pair production at hadronic colliders like LHC from their angular distribution in the rest frame of interaction. In contrast to this, we have also shown the rest frame angular distribution of spin-$\frac{1}{2}$ heavy fermions in type-$\mathrm{III}$ seesaw model considering the associated production of charged heavy neutral leptons along with their neutral counterpart. Once the spin determined, we carry forward our discussion with the probe of gauge representations of the leptoquarks of identical spins in electron-proton and electron-photon collisions by exploiting zeros in tree-level scattering amplitudes for the processes involving on-shell photons (Radiation Amplitude Zero, RAZ in short). We showed that the appearances of such zeros in these two colliders are complementary to each other and both the colliders are required for an exhaustive classification of the leptoquark models.
 
 This article is arranged as follows: the Leptoquark models have been discussed in Section~\ref{Sec:LQ}, Section~\ref{Sec:RAZ} gives a brief review of RAZ, different detectors considered for the leptoquark probes have been overviewed in Section~\ref{Sec:Detectors}, Leptoquark probes at hadronic collisions have been discussed in Section~\ref{Sec:LHC}, at electron-hadron collisions in Section~\ref{Sec:ep} and at electron-photon collisions in~\ref{Sec:eGam}. Finally we conclude this discussion with an outlook in Section~\ref{Sec:Concl}. 

%%%%%%%%%%%%%%%%%%%%%%%%%%%%%%%%%%%%%%%%%%%%%%%%%%%%%%%%%%%%%%%%%%%%%%%%%%%%%%%

\section{The Leptoquarks} \label{Sec:LQ}
The leptoquarks are proposed  bosons with spins-0 or 1 which carry lepton and baryon numbers simultaneously, and therefore any leptoquark shares  three-point vertex with a quark and a lepton \cite{Dorsner:2016wpm}. They appear naturally in  various extensions of SM with higher gauge groups, like $SO(10)$ in GUT scenarios, and they can address several issues like B-anomalies, muon $(g-2)$ anomaly, neutrino oscillations, etc. successfully. These hypothetical particles are all colour triplets, but they can have different $SU(2)_L$ representations. Table~\ref{Tab:LQ} shows different leptoquarks, their $SU(3)$ representations,weak hypercharge ($\mathtt Y_\phi$), respective projections ($T_3$) of $SU(2)_L$ representations, electromagnetic charges ($Q_{\phi}$) and their respective interaction lagrangians with SM fermions. Leptoquarks have been generically denoted by $\phi$ with suffix $s$ or $v$ to denote scalar and vector respectively. The subscript 1, 2 and 3 indicate their $SU(2)_L$ nature. The vector leptoquark candidates are assigned additional suffix $\mu$ to explicit their Lorentz index.

%%%%%%%%%%%%%%%%%%%%%%%%%%%%%%%%%%%%%Leptoquark Models%%%%%%%%%%%%%%%%%%%%%%%%%%%%%%%%%
\begin{table}[!htb]
\tbl{Quantum numbers of scalar and vector Leptoquarks. The $S_3^{adj}$ and $U_{3\mu}^{adj}$ denote the scalar and vector triplet Leptoquarks in adjoint representation \cite{Bandyopadhyay:2020jez}. \label{Tab:LQ}} { %
\renewcommand{\arraystretch}{0.95}
\begin{tabular}{|cccccc|cccccc|} \hline 	\\ [-3mm]
$\phi$ & $SU(3)$ & $\mathtt Y_\phi$ & $T_3$ & $Q_{\phi}$ & Interaction (+ h.c.) & $\phi$ & $SU(3)$ & $\mathtt Y_\phi$ & $T_3$ & $Q_{\phi}$ & Interaction (+ h.c.)  \\ [1mm] \hline \\ [-3mm]
\multicolumn{5}{l}{\textbf{Scalar Leptoquarks} $\bm{\phi_s}$} &  & \multicolumn{5}{l}{\textbf{Vector Leptoquarks} $\bm{\phi_v}$} & \\
\multirow{2}{*}{$S_1$} & \multirow{2}{*}{$\overline 3$} & \multirow{2}{*}{$\nicefrac{2}{3}$} & \multirow{2}{*}{0} & \multirow{2}{*}{$\nicefrac{1}{3}$} & $Y_L \,\xbar \Q_L^c \, \left( i\sigma^2 \, S_1\right ) \bold{L}_L$ & \multirow{2}{*}{$U_{1\mu}$} & \multirow{2}{*}{3} & \multirow{2}{*}{$\nicefrac{4}{3}$} & \multirow{2}{*}{0} & \multirow{2}{*}{$\nicefrac{2}{3}$} & $Y_L \, \xbar \Q_L \gamma^\mu \, U_{1\mu} \LL_L$ \\
 & & & & & $+ Y_R \, \xbar \uq^c_R \, S_1 \,\cl_R$ & & & & & & $+ Y_R \, \xbar \dq_R \gamma^\mu \, U_{1\mu}\cl_R$  \\ [3mm]
$\widetilde S_1$ & $\overline 3$ & $\nicefrac{8}{3}$ & 0 & $\nicefrac{4}{3}$  & $ Y_R\, \xbar{\dq}^c_R\, \widetilde{S}_1\,\cl_R $ & $\widetilde U_{1\mu}$ & 3 & $\nicefrac{10}{3}$ & 0 & $\nicefrac{5}{3}$ & $Y_R\,\xbar{\uq}_R\, \gamma^\mu \,\widetilde{U}_{1\mu}\,\cl_R $  \\ [3mm]
\multirow{2}{*}{$R_2$} & \multirow{2}{*}{3} & \multirow{2}{*}{$\nicefrac{7}{3}$} & $\nicefrac{1}{2}$ & $\nicefrac{5}{3}$ & $Y_L\, \xbar{\uq}_R \left(i\sigma^2 R_2\right)^T \LL_L$ & \multirow{2}{*}{$V_{2\mu}$} & \multirow{2}{*}{$\overline 3$} & \multirow{2}{*}{$\nicefrac{5}{3}$} & $\nicefrac{1}{2}$ & $\nicefrac{4}{3}$ & $Y_L\,\xbar{\dq}_R^c\, \gamma^\mu \left(i\sigma^2 V_{2\mu}\right)^T \LL_L$ \\
 & & & $\nicefrac{-1}{2}$ & $\nicefrac{2}{3}$ & $+ Y_R\, \xbar{\Q}_L\, R_2\, \cl_R$ & & & & $\nicefrac{-1}{2}$ & $\nicefrac{1}{3}$ & $+ Y_R\,\xbar{\Q}_L^c \,\gamma^\mu \left(i\sigma^2 V_{2\mu}\right)\cl_R$  \\ [3mm] 
\multirow{2}{*}{$\widetilde{R}_2$} & \multirow{2}{*}{3} & \multirow{2}{*}{\nicefrac{1}{3}} & $\nicefrac{1}{2}$ &\nicefrac{2}{3} & \multirow{2}{*}{$Y_L\,\xbar\dq_R\left(i\sigma^2\widetilde{R}_2\right)^T\LL_L$} & \multirow{2}{*}{$\widetilde V_{2\mu}$} & \multirow{3}{*}{$\overline 3$} & \multirow{2}{*}{$\nicefrac{-1}{3}$} & $\nicefrac{1}{2}$ & $\nicefrac{1}{3}$ & \multirow{2}{*}{$Y_L\,\xbar\uq_R^c\,\gamma^\mu\left(i\sigma^2\widetilde{V}_{2\mu}\right)^T\LL_L$} \\
 & & & $\nicefrac{-1}{2}$ & $\nicefrac{-1}{3}$ & & & & & $\nicefrac{-1}{2}$ & $\nicefrac{-2}{3}$ &  \\ [3mm]
\multirow{3}{*}{$\vec{S}_3$} & \multirow{3}{*}{$\overline 3$} & \multirow{3}{*}{$\nicefrac{2}{3}$} & 1 & $\nicefrac{4}{3}$ & \multirow{3}{*}{$ Y_L \, \xbar{\Q}_L^c \left( i\sigma^2 \, S_3^{adj} \right) \LL_L $} & \multirow{3}{*}{$\vec{U}_{3\mu}$} & \multirow{3}{*}{$ 3$} & \multirow{3}{*}{$\nicefrac{4}{3}$} & 1 & $\nicefrac{5}{3}$ & \multirow{3}{*}{$Y_L \,\xbar{\Q}_L \, \gamma^\mu \, U_{3\mu}^{adj} \, \LL_L $} \\
 & & & 0 & $\nicefrac{1}{3}$ & & & & & 0 & $\nicefrac{2}{3}$ &  \\
 & & & $-1$ & $\nicefrac{-2}{3}$ & & & & & $-1$ & $\nicefrac{-1}{3}$ &  \\ [1mm] 
\hline
\end{tabular} }	
\end{table}
%%%%%%%%%%%%%%%%%%%%%%%%%%%%%%%%%%%%%%%%%%%%%%%%%%%%%%%%%%%%%%%%%%%%%%%%%%%%%%%

\begin{figure}[!htb] 
\centering
\hspace*{-0.75cm}
\includegraphics[width=0.24\textwidth]{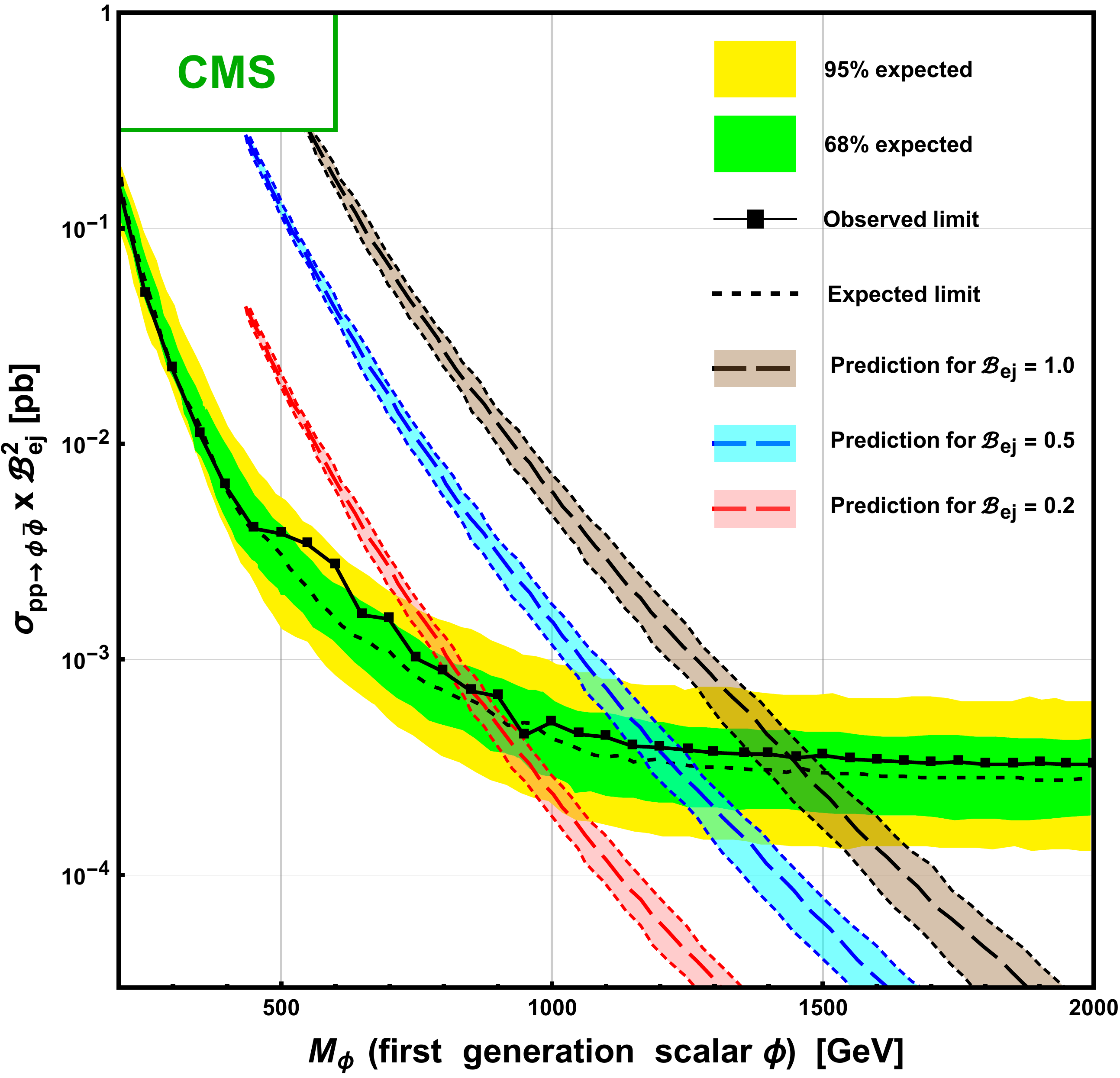}\hfil
\includegraphics[width=0.24\textwidth]{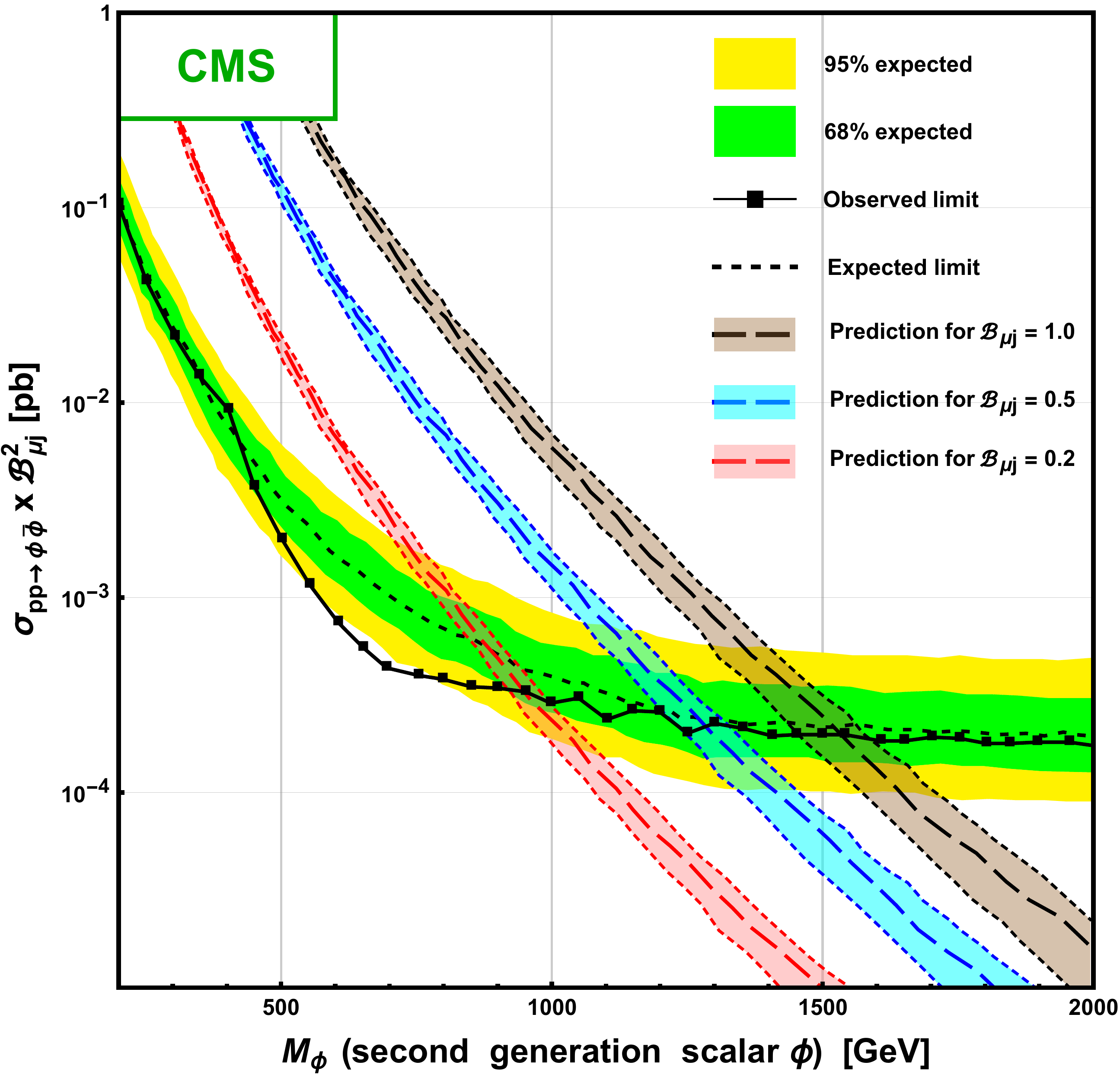}\hfil
\includegraphics[width=0.24\textwidth]{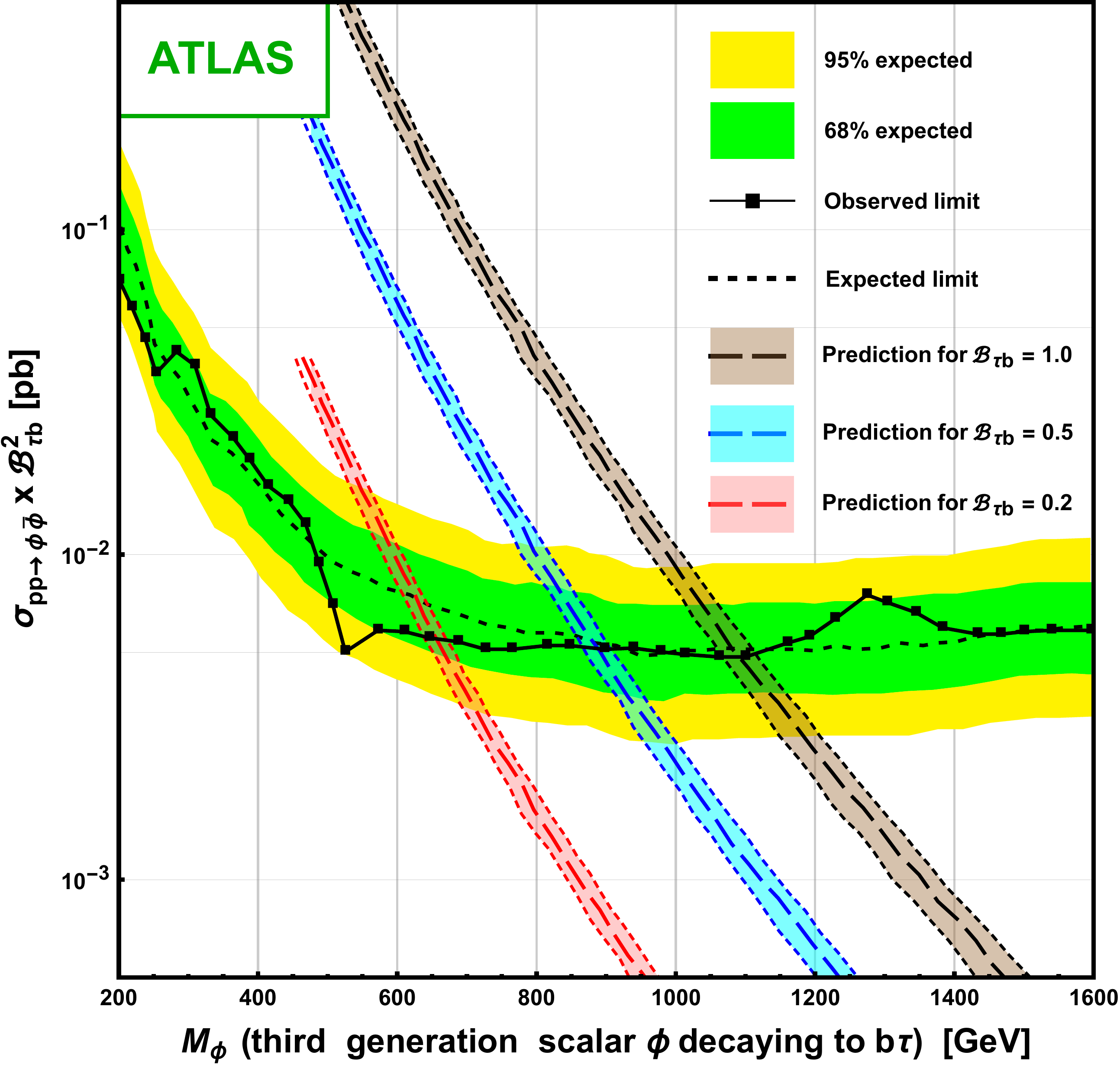}\hfil
\includegraphics[width=0.24\textwidth]{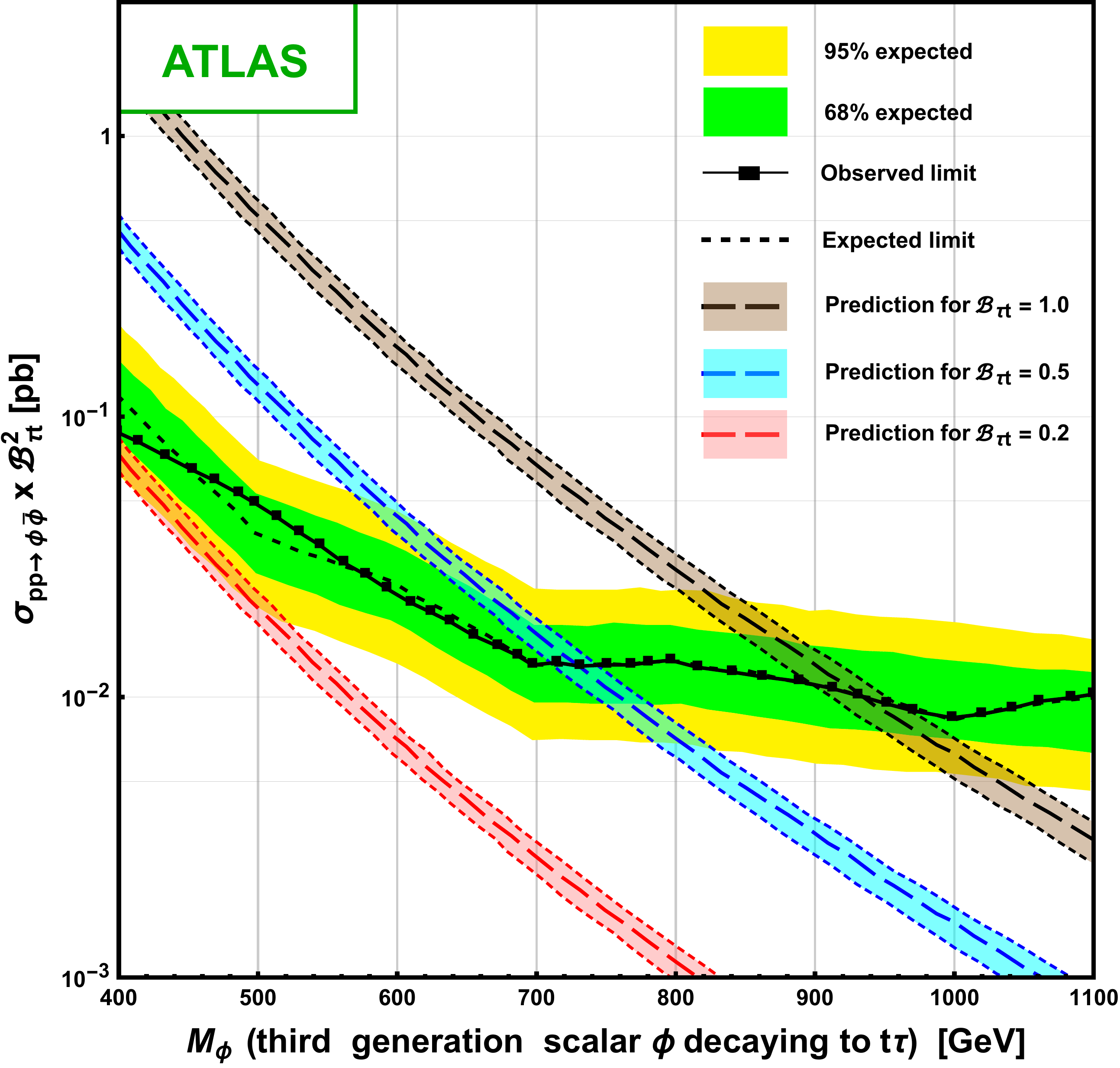}

\hspace*{-0.75cm}
\includegraphics[width=0.235\textwidth]{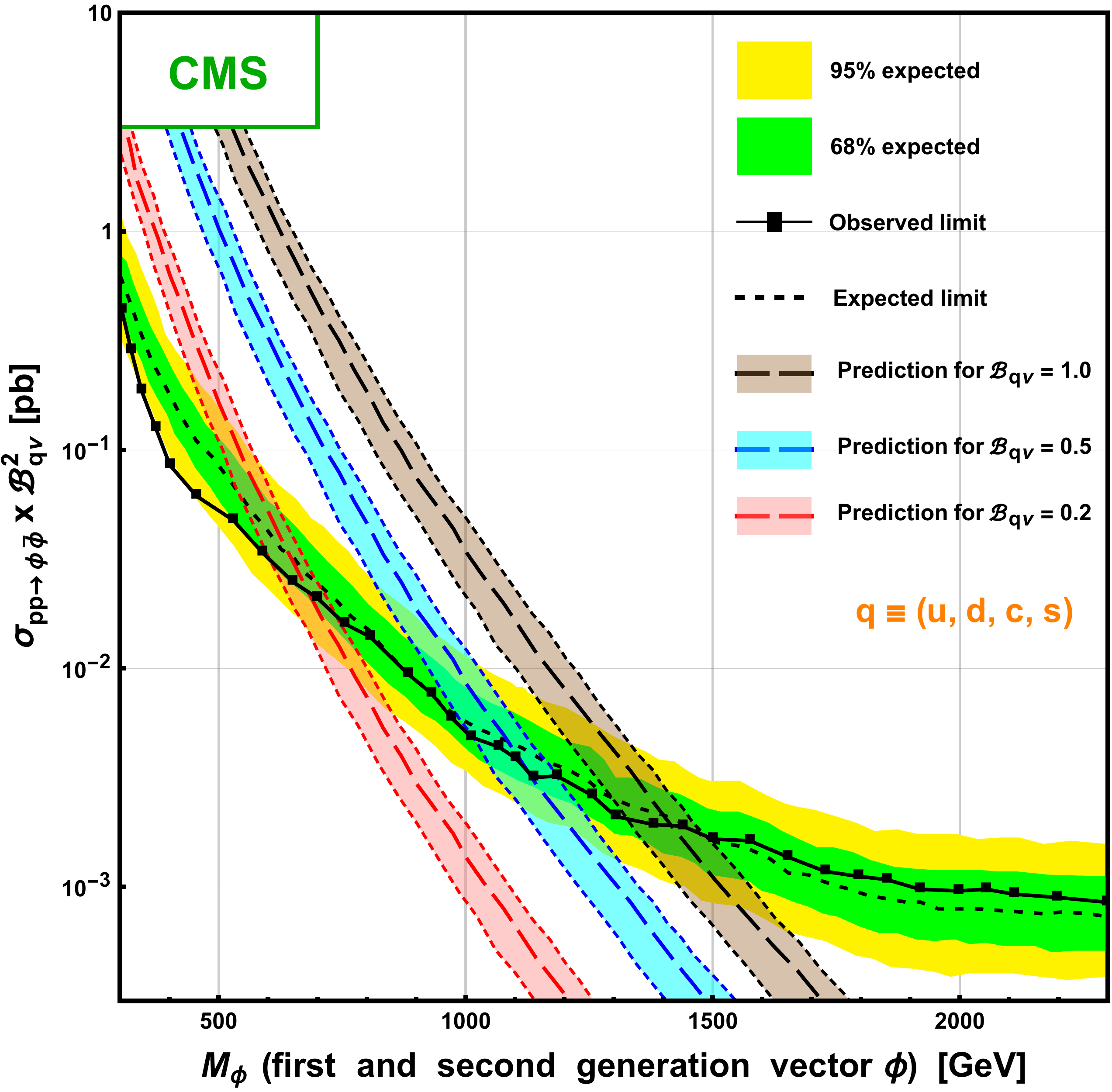}\hfil
\includegraphics[width=0.235\textwidth]{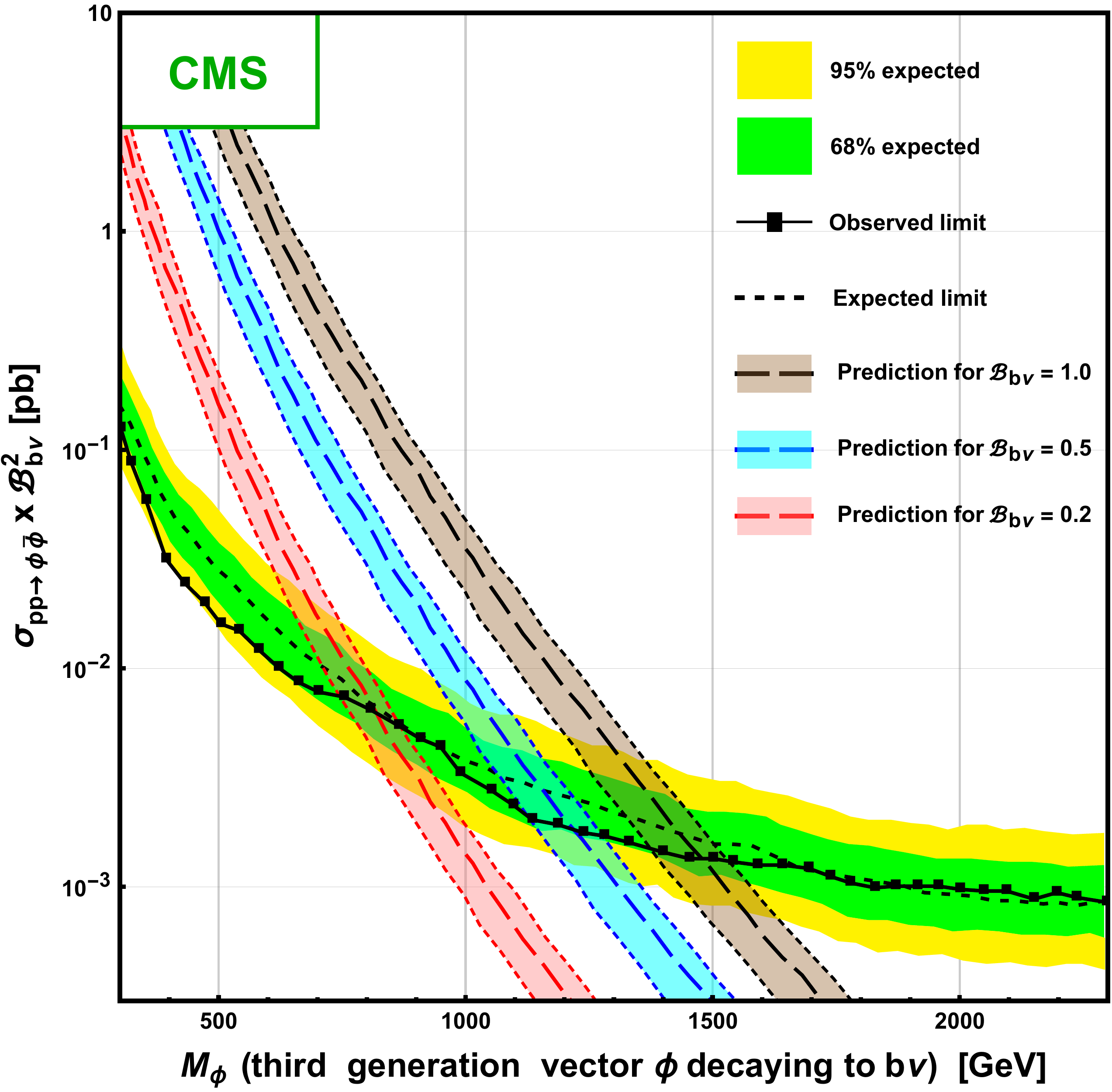}\hfil
\includegraphics[width=0.235\textwidth]{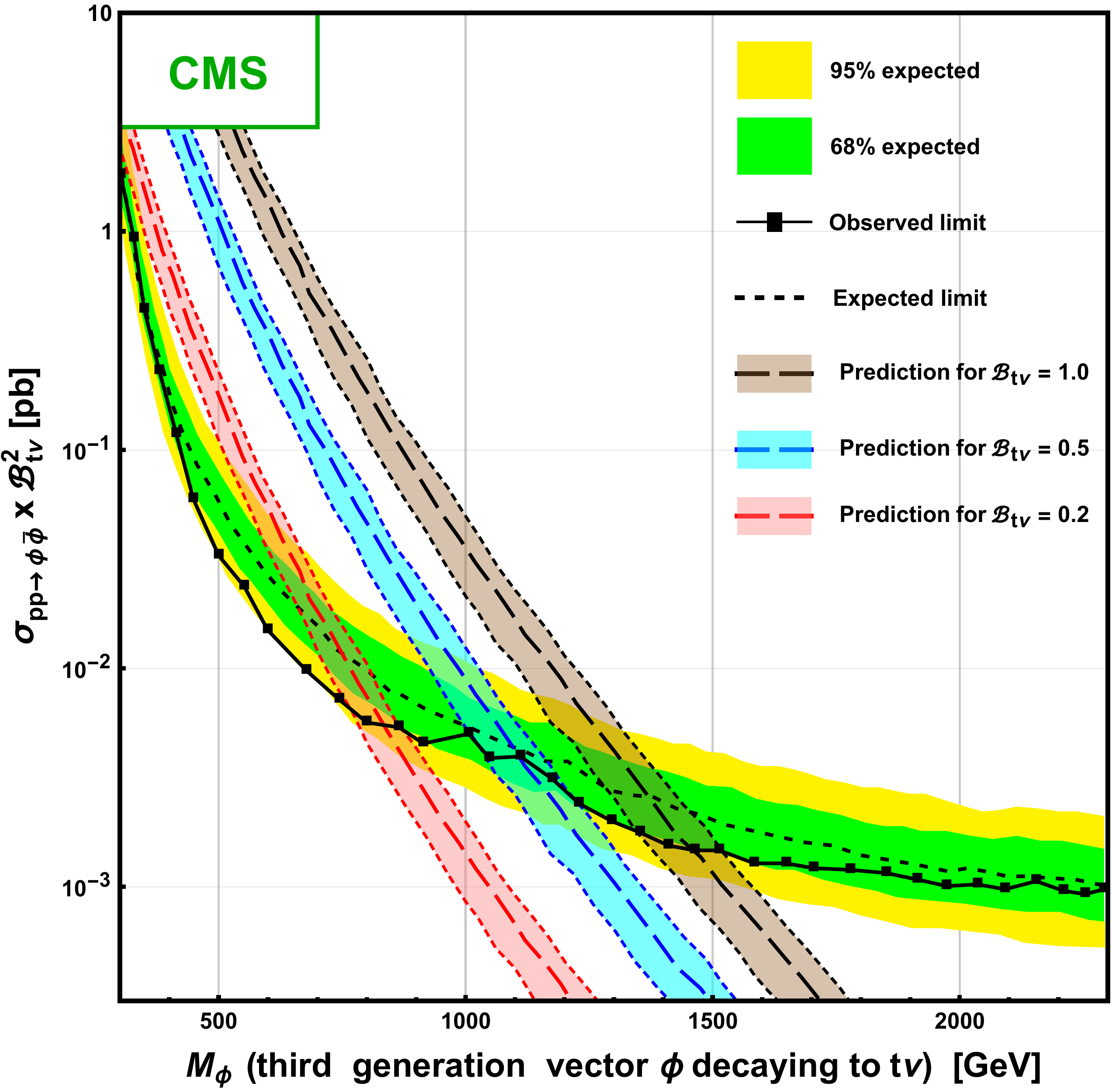} \hfil
\includegraphics[width=0.235\textwidth]{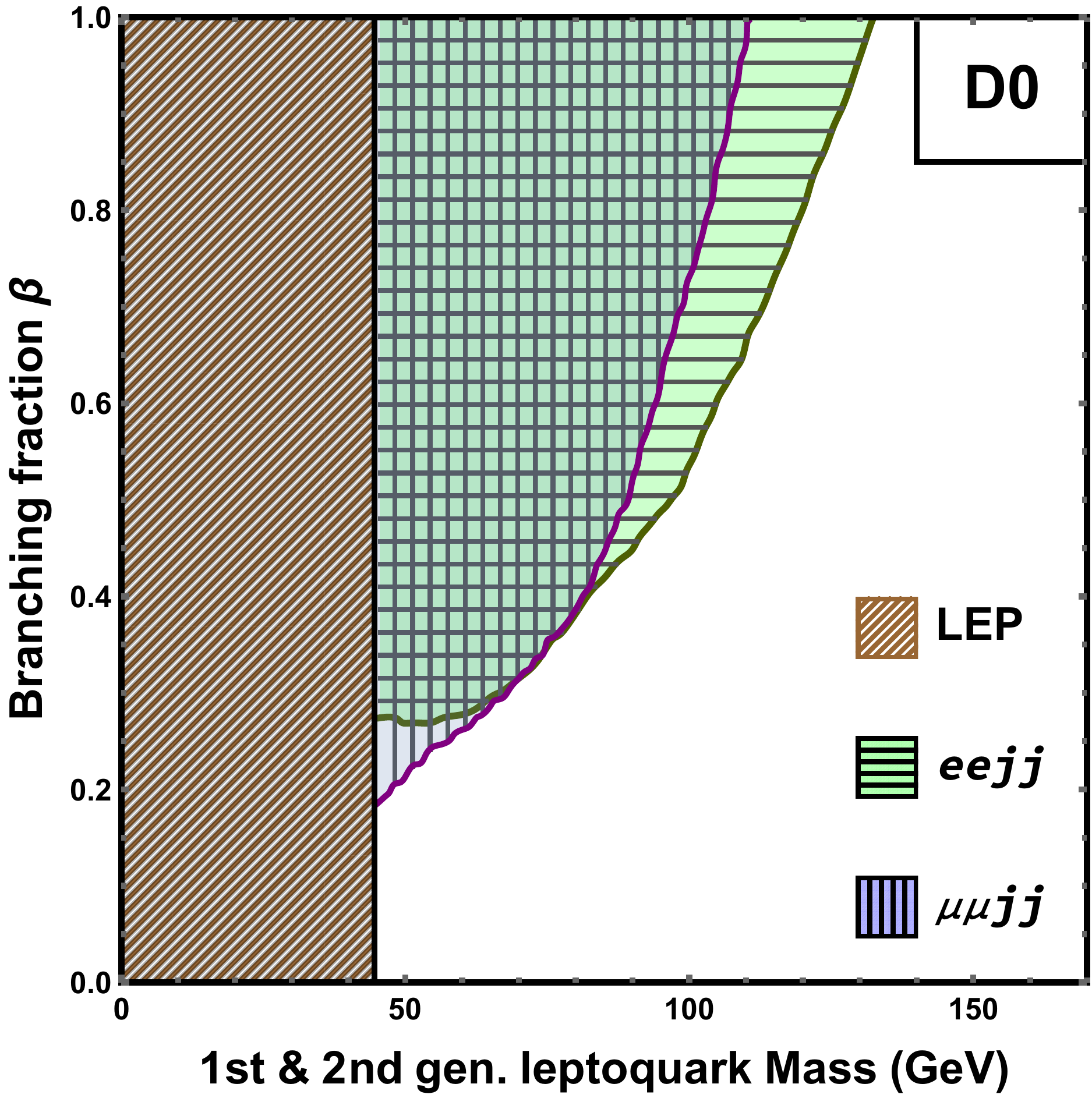}
 
\caption{Several bounds on scalar and vector leptoquarks. First row indicates the CMS and ATLAS constraints on different generations of scalar leptoquarks from visible decay modes \cite{Bandyopadhyay:2020jez}. The first three plots in the second row illustrates the CMS restrictions on different generations of vector leptoquarks from invisible decay modes\cite{Bandyopadhyay:2020jez}. The last plot in the second row signifies bounds from D$\slashed0$ and CDF while considering the light leptoquarks \cite{Bandyopadhyay:2020klr}.}
\label{Fig:LQBounds}
\end{figure}

Our choices for different leptoquark benchmark points are guided by the latest collider bounds provided by CMS and ATLAS collaborations \cite{Aaboud:2019bye,Aaboud:2019jcc,Sirunyan:2018btu,Sirunyan:2018ryt,Sirunyan:2018kzh,Sirunyan:2018nkj,Sirunyan:2018vhk}, as illustrated in Figure~\ref{Fig:LQBounds}. As the most recent analysis with CMS data suggests, masses of scalar leptoquarks decaying completely to first and second generation of charged leptons and quarks have been ruled out below 1435 and 1530 GeV respectively \cite{Sirunyan:2018btu,Sirunyan:2018ryt}. For the third generation scalar Leptoquarks , masses below 900 and 1020 GeV have been ruled out by 95\% confidence limit, if they decay respectively to $t\tau$ and $b\tau$  channels with 100\% branching \cite{Sirunyan:2018kzh,Sirunyan:2018nkj}. On the contrary, bounds on vector leptoquarks have been solely laid from their decays to neutrinos and quarks. CMS collaboration rules out vector leptoquarks with masses below 1115 GeV if they decay to $t\tau$ and $b\tau$ channels with 50\% branching in each, provided they couple with gluons minimally \cite{Sirunyan:2018vhk}. However these constraints drawn upon the assumption of leptoquark decays to a sole generation of lepton and quark loosen in case one considers leptoquark couples to all generations of quarks and leptons. However, when considering light mass leptoquarks we consider the D$\cancel{0}$ collaboration data at Fermilab, still uncontradicted by other observational imputs, which allows leptoquarks with mass around 70 GeV and $\sim 25\%$ branching to each of the first and second generation leptons and quarks \cite{CDF_D0}.

Considering all these bounds, we choose several benchmark points for our analysis. For leptoquark pair-production at LHC, we considered TeV-scale leptoquarks with masses 1.0 TeV, 1.5 TeV and 2.0 TeV respectively with diagonal Yukawa couplings to be 0.2 for all generations. For leptoquark probes at $ep$ collisions, we consider leptoquark masses of 70 GeV, 900 GeV, 1.5 TeV and 2.0 TeV with Yukawas complying with the above-mentioned bounds for each leptoquark models. Finally for $e\gamma$ collider, we consider leptoquarks with masses 70 GeV, 650 GeV and 1.5 TeV with appropriate couplings satisfying the bounds from direct and indirect probes.

The next section is devoted to a discussion of radiation amplitude zero, a notion exploited in leptoquark probes at $ep$ and $e\gamma$ colliders where a photon is present in either initial or final state of interaction.

%%%%%%%%%%%%%%%%%%%%%%%%%%%%%%%%%%%%%%%%%%%%%%%%%%%%%%%%%%%%%%%%%%%%%%%%%%%%%%%%%%%%%%%%%%%%%%%%%%%%%%%%%%%%%%%%%%%%%%%%%%%%%%%%%%%%%%%%%%%%%

\section{Zeros in Radiation Amplitudes (RAZ)} \label{Sec:RAZ}
Radiation Amplitude Zero was first observed in the context of the interaction, $q + \bar{q}^\prime \to W^\pm + \gamma$ where it was found 
that the distribution of the scattered photon (equivalently, the $W^\pm$) at the partonic rest frame of interaction vanishes at a certain angle with respect to the axis of the colliding beams \cite{Mikaelian,brown1}. Later Brown, Brodsky and Kowalsky showed that the tree-level amplitude with single photon of 4-momentum $k$ in a scattering process involving a total of $n$ initial and final particles of spins $\leq 1$ with charges $Q_i$ and 4-momenta $p_i$ vanishes, independent of the spins of interacting particles, at  certain kinematical zone characterised by identical ratios for $\frac{Q_i}{p_i \cdot k}$ for all, i = $1, 2, \cdots , n$ provided all the couplings are minimal \cite{brodsky,brown2}.

Now, for a generic $2 \to 2$ scattering process, $f_1 + f_2 \to f_3 + \gamma$,
the above condition reduces to
\begin{equation}
\cos\theta^* = \frac{Q_{f_1} - Q_{f2}}{Q_{f_1} + Q_{f2}}
\end{equation}
where $Q_{f_{1,2}}$ represents the electric charge of $f_{1,2}$ and $\theta^*$ denotes the angle between the photon and $f_1$ in rest frame of interaction at which the RAZ occurs. Here, masses of the initial particles have been neglected with respect to the energy of collision. As will be encountered in Section~\ref{Sec:eGam}, for the process
$e + q \to \phi + \gamma$ the above condition leads to,
\begin{equation} \label{Eq:RAZ1}
\cos\theta^* = 1 + \frac{2}{Q_\phi}
\end{equation}
where $Q_\phi$ is the electric charge carried by $\phi$. On the contrary, for the associated production of leptoquark with a quark at $e\gamma$ collision, i.e. $e + \gamma \to \phi + q$, the above-mentioned general condition now becomes,
\begin{equation} \label{Eq:RAZ2}
\cos\theta^* = 1 + \frac{2 Q_q}{1 - M_\phi^2/s}
\end{equation}
where, $\theta^*$ is the angle made by the photon and the quark in the centre of momentum (CM) frame , $Q_q$ the electric charge of $q$, $M_\phi$ is the mass of the leptoquark and $s$ the square of collision energy at CM frame.

It is evident from the Eqns.~\ref{Eq:RAZ1} and~\ref{Eq:RAZ2} that, in order to observe a zero in the angular distribution of the scattered photon produced in association with a leptoquark at $ep$ collision, the leptoquark must have $\lvert Q_\phi \rvert > 1$
whereas, zero in the angular distribution of the scattered quark produced in association with a leptoquark at $e\gamma$ collision appears only when $\lvert Q_\phi \rvert < 1$. This establishes the complementarity in functionality of the two colliders in relation to an exhaustive analysis of all the leptoquark candidates. 

\begin{figure}[!htb]
	\centering
	\mbox {
		\subfigure[$Q_q = -1/3$]{\includegraphics[width=0.3\textwidth]{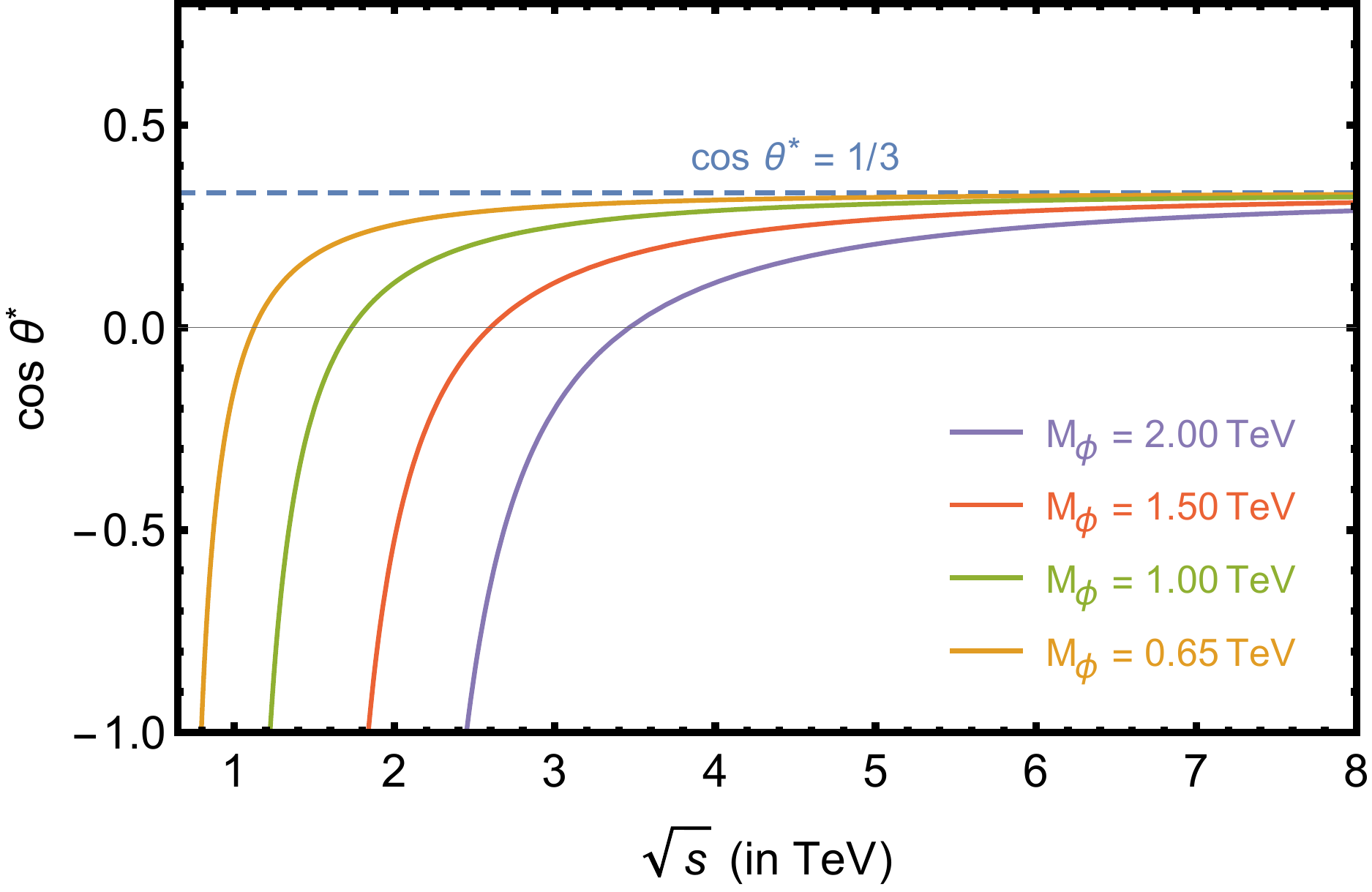}}
		\hspace*{2.0cm}
		\subfigure[$Q_{\bar{q}} = -2/3$]{\includegraphics[width=0.3\textwidth]{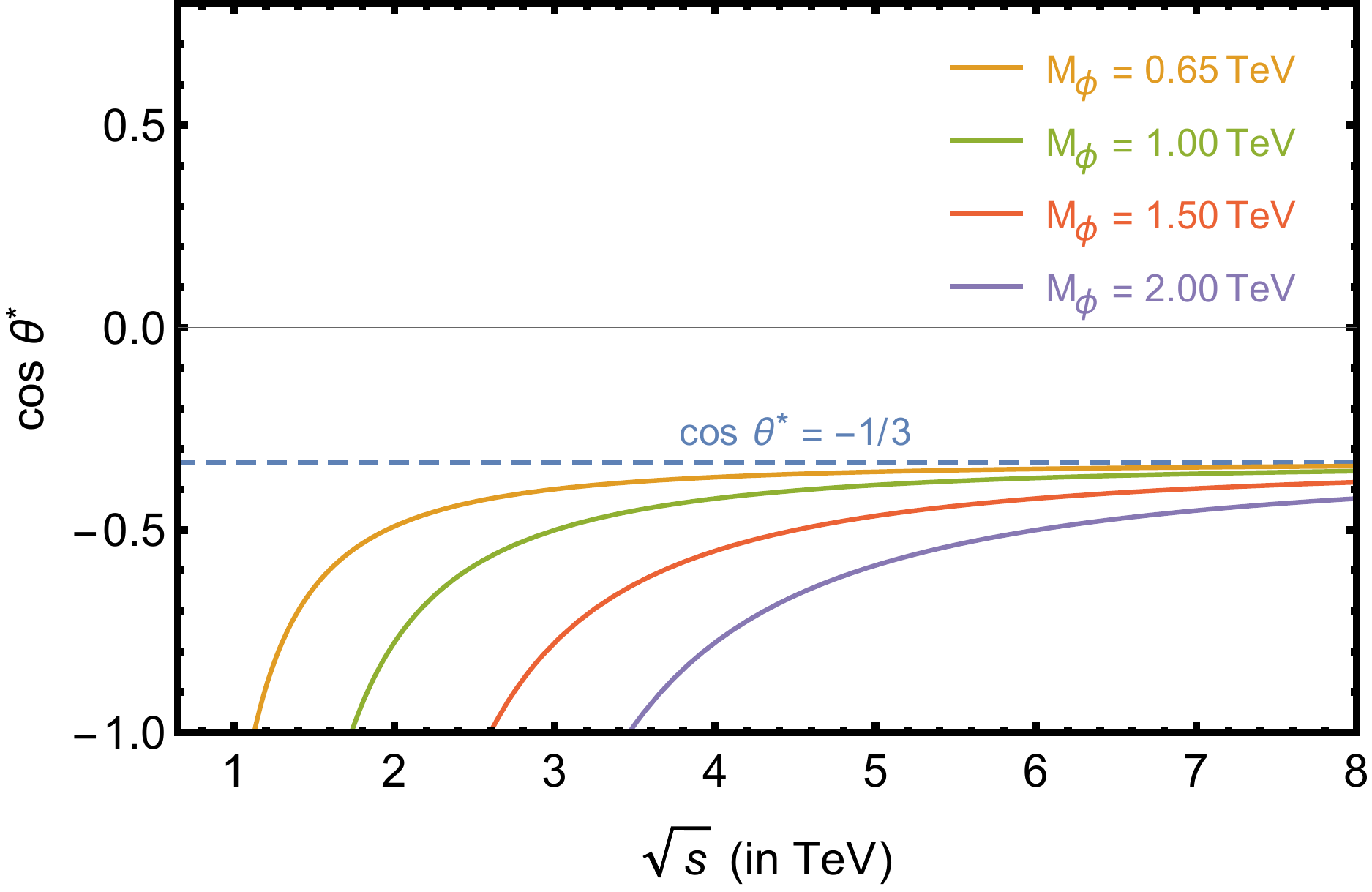}} }
	\caption{Variation of $\cos\theta^*$  with respect to the energy of interaction at CM frame for different leptoquark masses \cite{Bandyopadhyay:2020klr}.} \label{Fig:RAZEGamCos}
\end{figure}

We also note the dependence of the zero in scattering angle on the leptoquark mass $M_\phi$ and energy of interaction $\sqrt{s}$ in the rest frame for $e\gamma$ collision while the position of zero is independent of these factors in $ep$ collision. Figure~\ref{Fig:RAZEGamCos} shows the variation of the angle made by the quark with the photon in the rest frame of interaction along which no scattering of the quark occurs. As evident from the Eqn.~\ref{Eq:RAZ2}, for a given leptoquark mass this angle asymptotically converges to $\cos\theta^* = \pm 1/3$ depending on the charge of the quark (or equivalently the leptoquark).

The next section presents an overview of different collision parameters considered for our leptoquark probes. 

%%%%%%%%%%%%%%%%%%%%%%%%%%%%%%%%%%%%%%%%%%%%%%%%%%%%%%%%%%%%%%%%%%%%%%%%%%%%%%%%%%%%%%%%%%%%%%%%%%%%%%%%%%%%%%%%%%%

\section{The Detectors} \label{Sec:Detectors}
We begin our analysis with the hadronic collisions. For the purpose of our study, we consider the $pp$ collisions at LHC at the centre of mass energies ($\sqrt{s}$) of 14, 30 and 100 TeV respectively with an integrated luminosity of 1000 fb$^{-1}$. This can be achieved in the future upgrades at LHC or FCC \cite{Abada:2019ono,BejarAlonso:2020kmn}.

We next consider the leptoquark probes in $ep$ colliders, namely at Run-6 of the Large Hadron Electron Collider (LHeC) \cite{Agostini:2020fmq,Bordry:2018gri} with asymmetric beam of electrons, $E_e = 50$ GeV and protons, $E_p = 7$ TeV, amounting to a collision energy of $\sqrt{s} = 1183.2$ GeV. We supplemented our searches with the Future Electron Hadron Circular Collider (FCC-he) \cite{Agostini:2020fmq,Bordry:2018gri} at Run-$\mathrm{I}$ with $E_e = 60$ GeV, $E_p = 20$ TeV, $\sqrt{s} = 2190.2$ GeV and at Run-$\mathrm{II}$ with $E_e = 60$ GeV, $E_p = 50$ TeV, $\sqrt{s} = 3464.1$ GeV. For each of these collisions, we analysed the signals for an integrated luminosity of 2000 fb$^{-1}$.

We finally consider the associated leptoquark production in $e\gamma$ collisions \cite{Ginzburg:1981vm,Velasco:2002vg,Badelek:2001xb}. We took three separate sources for photons, namely monochromatic photons, photons from laser backscattering \cite{Ginzburg:1981vm} and photons from Equivalent Photon Approximation (EPA) \cite{vonWeizsacker:1934nji,Budnev:1974de}. We consider symmetric beams of electrons and photons (or positrons for EPA) at the collision energies of 200 GeV, 2 TeV and 3 TeV, with an integrated luminosity of 100 fb$^{-1}$ for each case.

%%%%%%%%%%%%%%%%%%%%%%%%%%%%%%%%%%%%%%%%%%%%%%%%%%%%%%%%%%%%%%%%%%%%%%%%%%%%%%%%%%%%%%%%%%%%%%%%%%%%%%%%%%%%%%%%%%

\section{Detector Probes of Leptoquarks}
We perform a PYTHIA based simulation of Leptoquark probes in all three kinds of collisions detailed in the previous section. We detail the workflow in the following:

\begin{itemize}

\item The leptoquark models are separately written on SARAH, which are then executed to generate model files for CalCHEP. The model files are then used to generate events with the beam specifications evoked in Section~\ref{Sec:Detectors}. For protons, NNPDF2.3 has been used to account for the parton distribution function. The events are then stored in ".lhe" format. The ".lhe" event files are then read in PYTHIA8 where the radiation effects, parton showering and hadronisations are accounted for. Fastjet-3.2.3 has been used for jet reconstruction using anti-kT algorithm with a jet-radius $\Delta R$ of 0.5, from stable hadrons and photons appearing from $\pi^0$ decays.

\item The calorimeter coverage is taken to be $\lvert \eta \rvert < 4.5, ~2.5$ respectively for the jets and leptons (or photons). For $ep$ collisions due to high asymmetry of the colliding beams, the calorimeter coverage for leptons (appearing from leptoquark decays) and photons has been extended to 4.5.

\item The tagged leptons (and photons) are clean of hadronic activities which implies that the hadronic activity within a cone of $\Delta R < 0.3 (0.2)$ around each lepton (and photon), is less than 15\% of the leptonic (photonic) transverse momentum ($p_T$). 

\item The minimum $p_T$ for the jets, leptons and photons have been demanded to be 20 GeV. Leptons are well separated from jets $\Delta R_{lj} > 0.4$ and other leptons $\Delta R_{ll} > 0.2$. The same hold for the accepted photons, $\Delta R_{\gamma j} > 0.2$ and $\Delta R_{\gamma l} > 0.2$.

\end{itemize}

%%%%%%%%%%%%%%%%%%%%%%%%%%%%%%%%%%%%%%%%%%%%%%%%%%%%%%%%%%%%%%%%%%%%%%%%%%%%%%%%%%%%%%%%%%%%%%%%%%%%%%%%%%%%%%%%

\subsection{Leptoquark Spins at Hadronic Collider} \label{Sec:LHC}
As an illustration of segregation of leptoquarks based on spins, in pair-production at $pp$ collisions we consider the scalar singlet leptoquark $S_1^{1/3}$ along with the vector singlet $\widetilde{U}_{1\mu}^{5/3}$ \cite{Bandyopadhyay:2020wfv}. The pair-production cross-section is QCD dominated and hence more or less same for all the leptoquarks with same spins.

We consider all dominant, irreducible SM backgrounds, namely, $t\bar{t}$, $t\bar{t}W^\pm$, $t\bar{t}Z$, $tW^-Z$, $\bar{t}W^+Z$, $W^+W^-$, $W^\pm Z$, $ZZ$, $W^+W^-W^\pm$, $W^+W^-Z$, $ZZW^\pm$, $ZZZ$. In order to optimise signal events over backgrounds we only accept the muons and jets from the leptoquark decay. Hence, we analyse the channels with $S_1^{1/3} \to \mu^+ \bar{c}$ and $\widetilde{U}_{1\mu}^{5/3} \to \mu^+ c$. We therefore demand events with $\geq 1\mu^+ + 1\mu^- + 2j$. In order to reject backgrounds involving an on-shell Z boson, we impose every combination of opposite charged leptons and jets to satisfy $\lvert M_{\ell\ell} - M_Z \rvert > 5$ and $\lvert M_{jj} - M_Z \rvert > 10$ GeV. We then consider all possible combinations of the jet-lepton pairs  to evaluate their invariant mass. The pairs originated from the leptoquark decay will peak at the invariant leptoquark mass while the rest will form a continuum. However, for the SM backgrounds, the pattern will show a monotonic fall with the increase in the jet-lepton invariant mass. We finally select signal events demanding exactly 1 $\mu^- j$ and 1 $\mu^+ j$ invariant masses falling within a 10 GeV window around these leptoquark resonance peaks. We also study the kinematics of the leptoquark decay products from the $p_T$s of these jets ($js$), muons and antimuons ($\mu s$) at $\sqrt{s} = 14$ TeV. We observe that the jet and muon $p_T$s peak roughly around half the leptoquark mass, independent of the leptoquark spin. We note the presence of longer tails for jet and muon $p_T$s for 100 TeV collisions, compared to the 14 TeV counterpart.

We now analyse the angular distribution of the scattered leptoquark pair, completely reconstructed from respective decay products in the rest frame of interaction, evaluated from the 4-momenta of the reconstructed pairs. The partonic angular distribution depends on its spin as well as nature of Yukawa coupling it possesses with lepton and quark. However, since the pair-production is QCD dominated, effects of Yukawa couplings get suppressed in real collider. Moreover, the interacting partons carry a fraction of colliding proton 4-momenta with probabilities given by the parton distribution functions. Due to these effects, the actual angular distribution observed in $pp$ collider will be a bit different from the partonic angular distribution. Nevertheless, the distribution pattern is unique to the leptouark spin. For scalar leptoquarks the distribution takes a convex shape while the same for any vector leptoquark becomes concave.

%\end{figure}

\begin{figure}[!htb]
\centering
\hspace*{-1.0cm}
\mbox {
\subfigure[$\sqrt{s} = 14$ TeV]{\includegraphics[width=0.3\textwidth]{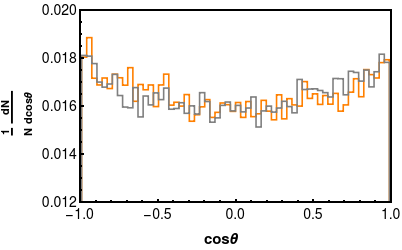}}
\subfigure[$\sqrt{s} = 30$ TeV]{\includegraphics[width=0.3\textwidth]{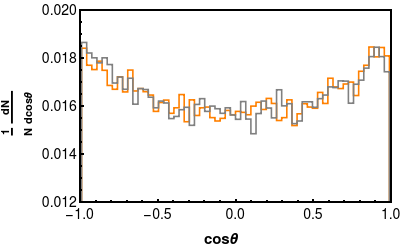}}
\subfigure[$\sqrt{s} = 100$ TeV]{\includegraphics[width=0.34\textwidth]{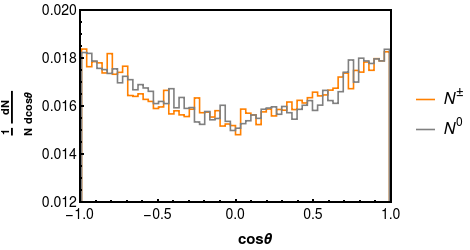}} }
\caption{Normalised angular distribution of 1.5 TeV Right Handed Neutrinos (RHNs) for $pp \to N^\pm N^0$ at the rest frame of interaction. We consider $N^\pm \to Z \ell^\pm$, $Z \to \ell^+ \ell^-$ and $N^0 \to W^\pm \ell^\mp$ and $W^\pm \to q \bar{q}^\prime$. Events with $\geq 4\ell + 2j$ are considered, same flavour opposite sign lepton pair invariant mass ($M_{\ell\ell}$) reconstructed along with jet-pair invariant mass ($M_{jj}$). Events with exactly 1 $\lvert M_{\ell\ell} - M_Z \rvert < 5$ GeV along with exactly 1 $\lvert M_{jj} - M_W \rvert < 10$ GeV has been then taken. From them, events with exactly 1 $\lvert M_{\ell\ell\ell} - M_N \rvert < 10$ GeV and 1 $\lvert M_{jj\ell} - M_N \rvert < 10$ GeV have been finally taken.} \label{Fig:RHN}
\end{figure}

As an additional example, we study the angular distribution for the pair production of spin-$\frac{1}{2}$ heavy leptons in Type-$\mathrm{III}$ seesaw scenario \cite{Foot:1988aq}. In Figure~\ref{Fig:RHN}, we present the angular distribution of scattered  heavy leptons with identical mass, at identical collision energies. We consider for the process $pp \to N^0 N^\pm$ with right handed neutrino decays to charged leptons and jets such that they are perfectly reconstructible. The result reinforce our claim: angular distribution of the scattered states, pair-produced in a $2 \to 2$ scattering, in rest frame of interaction bears unique pattern based on the spin of the scattered states.

It is also important to mention that for any given energy of collision and mass of leptoquark, the vector leptoquarks have much higher production cross-section compared to their scalar counterparts since vectors have three spin degrees of freedom which enhance the cross-section for pair production by a factor of nine compared to the scalar one. Therefore, very less integrated luminosity is required to produce the vector leptoquarks. Thus the vector leptoquarks can be discovered or ruled out at very early stage of any high energetic $pp$ collider.

Now, in order to distinguish different members of the same $SU(2)_L$ multiplet,
 jet charge can be an effective observable, considering leptoquark decays to a certain generation of charged lepton and quark. Let us consider the vector doublet $V_{2\mu}$, with components $V_{2\mu}^{4/3}$ and $V_{2\mu}^{1/3}$. Considering their decays to second generation charged leptons, we observe $V_{2\mu}^{4/3} \to \mu^+ \bar{s}$ while $V_{2\mu}^{1/3} \to \mu^+ \bar{c}$. Hence for a given leptoquark mass, if we look at the electromagnetic charge of jets appearing along with $\mu^+$ from decays of $V_{2\mu}^{4/3}$ and $V_{2\mu}^{1/3}$ respectively, we will observe that the former peaks roughly around 0.35 while the latter around -0.4. Therefore an appropriate cut over the charge of the jet from the leptoquark decay could optimise one member over the other in the same multiplet.

%%%%%%%%%%%%%%%%%%%%%%%%%%%%%%%%%%%%%%%%%%%%%%%%%%%%%%%%%%%%%%%%%%%%%%%%%%%%%%%%%%%%%%%%%%%%%%%%%%%%%%%%%%%%%%%%%%%%%%%%%%

\subsection{Electromagnetic Charge Probes of Leptoquarks in Electron Hadron Collisions} \label{Sec:ep}

Now we focus on differentiating the leptoquarks according to their electromagnetic charges through RAZ at $ep$ collider \cite{Bandyopadhyay:2020jez}. It would also help in separating different excitations of same $SU(2)_L$ multiplet from each others. As described in Eqn.~\ref{Eq:RAZ1}, leptoquarks with electromagnetic charge $-4/3$ (i.e. charge conjugates of $\widetilde S_1, S_3^{4/3}$ and $V_{2\mu}^{4/3}$) produced in association with a photon in $ep$ collision would exhibit zero at $\cos\theta = -0.5$ whereas the leptoquarks with charge $-5/3$ (i.e. charge conjugates of $R_2^{5/3}, \widetilde{U}_{1\mu}$ and $U_{3\mu}^{5/3}$) would exhibit the same at $\cos\theta = -0.2$ while observing angular distribution in the rest frame of interaction with $\theta$ being the angle between the photon and electron.

\begin{figure}[!htb]
	\centering
	\includegraphics[width=0.24\textwidth,height=0.11\textheight]{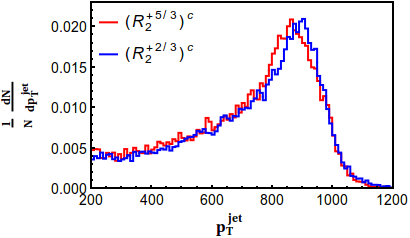}\hfil
	\includegraphics[width=0.24\textwidth,height=0.11\textheight]{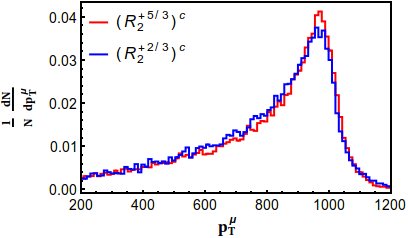}\hfil
	\includegraphics[width=0.24\textwidth,height=0.11\textheight]{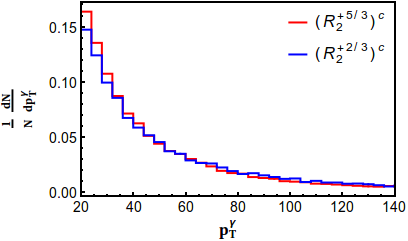}\hfil
	\includegraphics[width=0.24\textwidth,height=0.11\textheight]{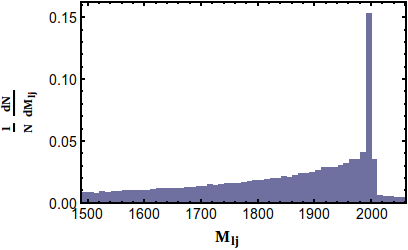} 
	\caption{Kinematic distributions at FCC II \cite{Bandyopadhyay:2020jez}. While the first three plots display the transverse momentum distributions for jets, muons and photons respectively, the last one shows invariant mass of all possible combinations of $\mu^- j$ for the signal events which peaks at $M_\phi=2000$ GeV.}
	\label{Fig:FCC2Mlj2}
\end{figure}

 We simulate the leptoquark production associated with a photon at $ep$ collider by PYTHIA8 for FCC where 60 GeV electrons will be collided with 20 TeV and 50 TeV protons 
 in run I and II respectively at an integrated luminosity of 2000 fb$^{-1}$. We take all the allowed couplings of leptoquarks with each generation of quark and lepton to be 0.2. For FCC run I, we simulated 1500 GeV leptoquarks, whereas for run II we analysed 2000 GeV leptoquarks. Here, we present the results for FCC run II only. In order to completely eliminate the SM backgrounds, we consider the  muon and the associated second generation quark decay mode solely. As mentioned in Sec.~\ref{Sec:RAZ}, the associated production of leptoquarks with magnitude of electromagnetic charge greater than unit electron charge would only exhibit zeros in the rest frame of interaction. We therefore demand each event with $\geq 1\mu^- + 1j + 1\gamma$ with $p_T$ $\geq 20$ GeV. We take all possible combinations of $\mu j$ pair and evaluate the invariant mass, which peaks at $M_\phi$. We then demand events with exactly one such combination within the 10 GeV window around the peak and exactly one hard photon with $p_T \geq 20$ GeV. The $p_T$ distributions of jets, muons and photons as well as the invariant distribution of muon-jet pair are displayed in Figure \ref{Fig:FCC2Mlj2}. We now determine the rest frame of interaction from the 4-momenta of the reconstructed leptoquark and hard photon. The angular distribution of the photon is finally evaluated in this boosted back frame of interaction.

\begin{figure}[!htb]
	\centering
	\mbox {
		\subfigure[$\widetilde{S}_1$]{\includegraphics[width=0.27\textwidth]{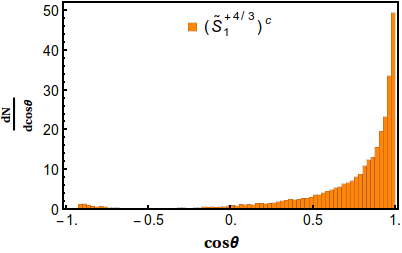}}
		\subfigure[$R_2$]{\includegraphics[width=0.27\textwidth]{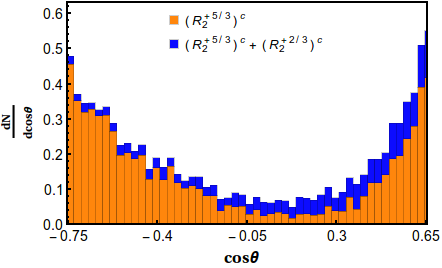}}  
		\subfigure[$S_3$]{\includegraphics[width=0.25\textwidth]{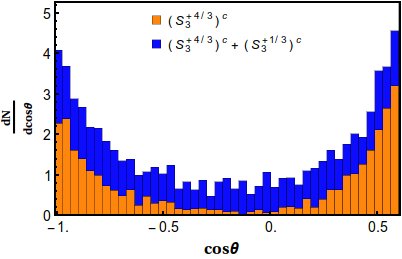}}  }
	
	\mbox {
		\subfigure[$\widetilde{U}_{1\mu}$]{\includegraphics[width=0.27\textwidth]{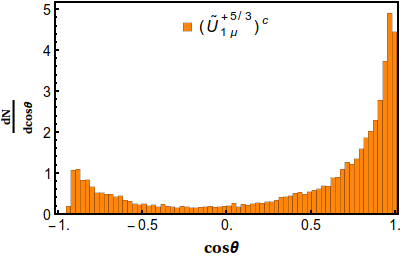}}
		\subfigure[$V_{2\mu}$]{\includegraphics[width=0.27\textwidth]{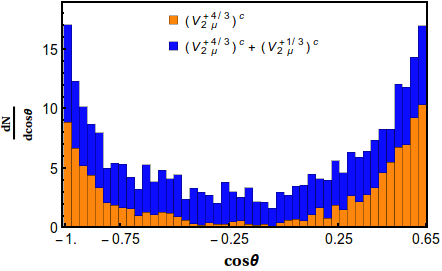}}  
		\subfigure[$U_{3\mu}$]{\includegraphics[width=0.26\textwidth]{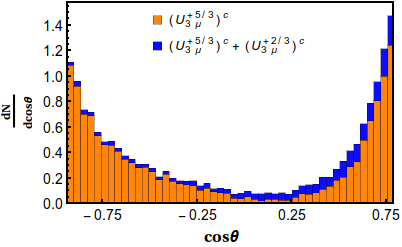}}  }
	\caption{Angular distribution for $ep\to\phi\gamma$ relative to angle between photon and electron beam at FCC-II with $\sqrt{s} = 3464.1$ GeV and $\mathcal{L}_{int} = 2000$ fb$^{-1}$ \cite{Bandyopadhyay:2020jez}. Different subfigures indicate the distributions for different leptoquarks.} \label{Fig:AngDisFCC}
\end{figure}
 
  In order to distinguish signature of a member of an $SU(2)_L$ multiplet showing RAZ from that of the other member in the multiplet which effaces the zero in the scattering amplitude, we exploit the possibility of imposing cut on electromagnetic charge of the jet originating from the leptoquark decay. As an illustration, let us consider the scalar doublet $R_2$. $R_2^{-5/3}$ would exhibit zero at $\cos\theta = -0.2$ whereas the distribution associated to the production of its degenerate partner, $R_2^{-2/3}$ would erase out this dip in scattering angle. In order to optimise the signature of $R_2^{-5/3}$ over $R_2^{-2/3}$, we impose an additional cut on the jet charge ($Q_J$) evaluated upon the jets appearing from the leptoquark decay, \textit{i.e,} with $\lvert M_{\mu j} - M_\phi \rvert \leq 10$ GeV. We considered only those events with $Q_J < 0.3$ to reject $R_2^{-2/3}$ signal substantially without significantly diminishing the $R_2^{-5/3}$ signal. It is interesting to mention that all the leptoquarks show significances more than $5\sigma$ at FCC run II with integrated luminosity of 2000 fb$^{-1}$.

 We show our results obtained for FCC-II in Figure~\ref{Fig:AngDisFCC}. The upper row shows the angular distribution of the scalars while the lower exhibits that for the vector leptoquarks. As apparent from these plots, a well distinguished minimum around the expected value of $\cos\theta$ exists for the signature of every multiplet.  
 
%%%%%%%%%%%%%%%%%%%%%%%%%%%%%%%%%%%%%%%%%%%%%%%%%%%%%%%%%%%%%%%%%%%%%%%%%%%%%%%%%%%%%%%%%%%%%%%%%%%%%%%%%%%%%5

\subsection{Complentary Probes at Electron Photon Collider} \label{Sec:eGam}

\begin{figure}[!htb]
	\centering
	\includegraphics[height=0.3\textwidth, width=0.15\textheight,angle=-90]{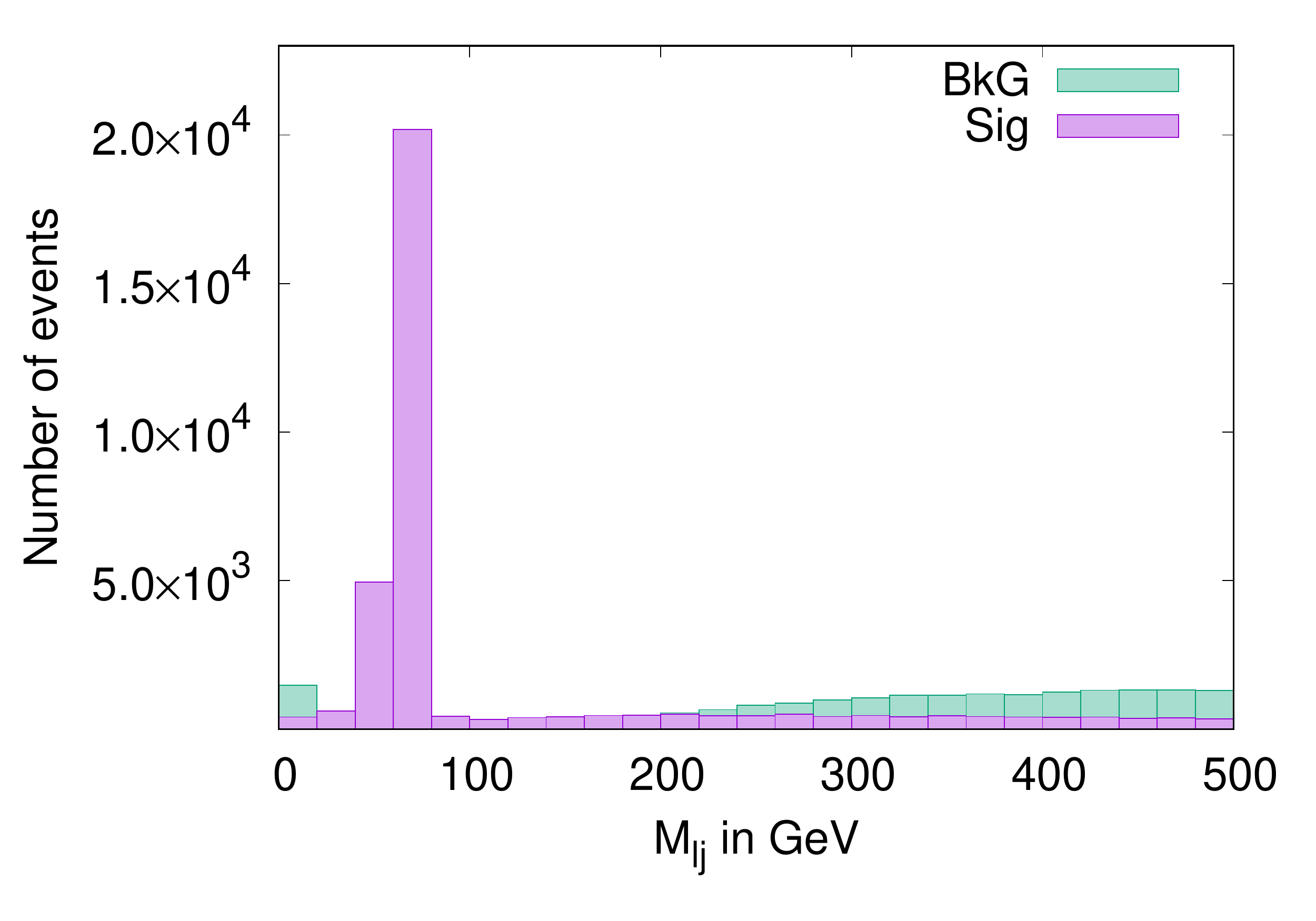}
	\includegraphics[height=0.3\textwidth, width=0.15\textheight,angle=-90]{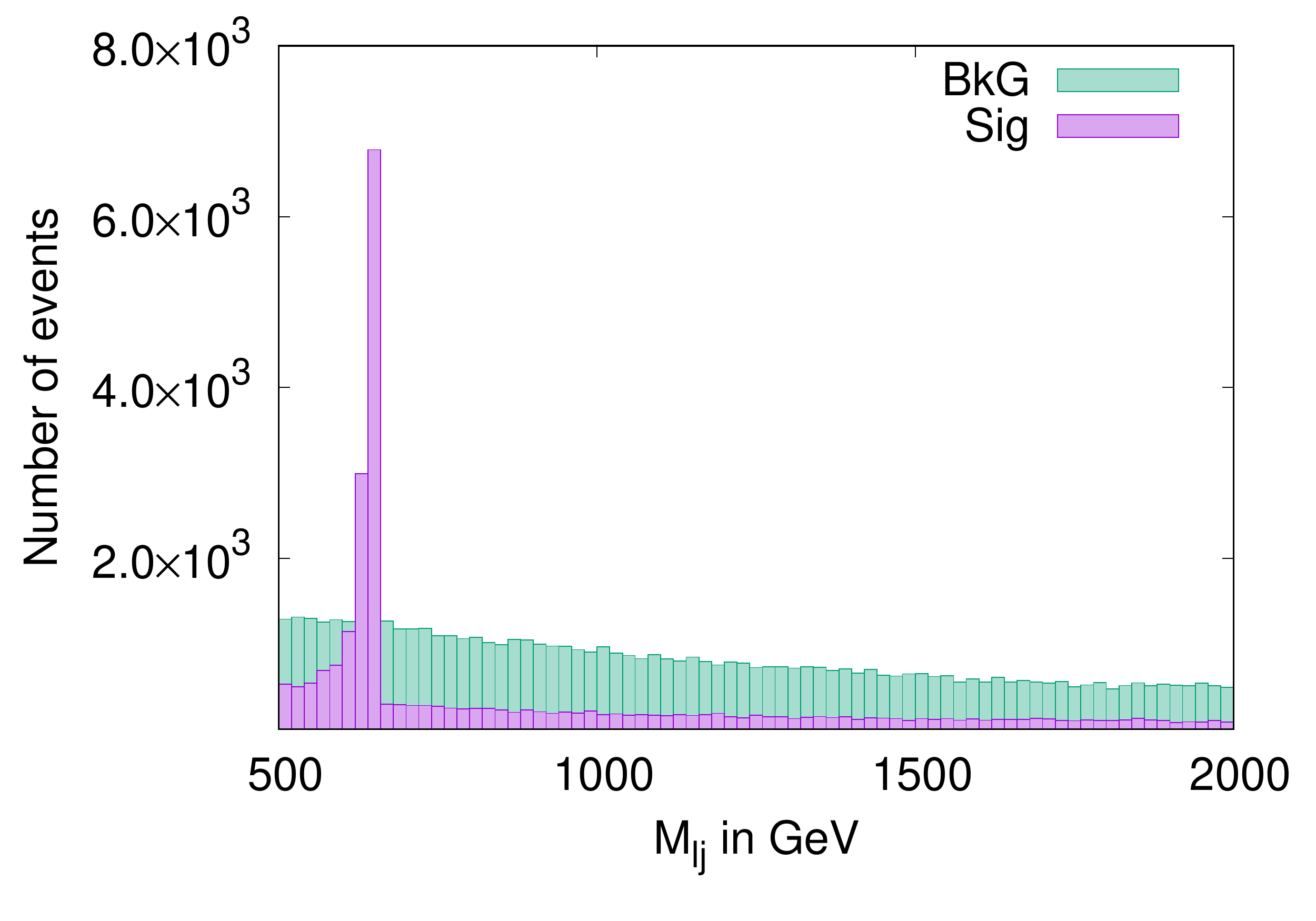}
	\includegraphics[height=0.3\textwidth, width=0.15\textheight,angle=-90]{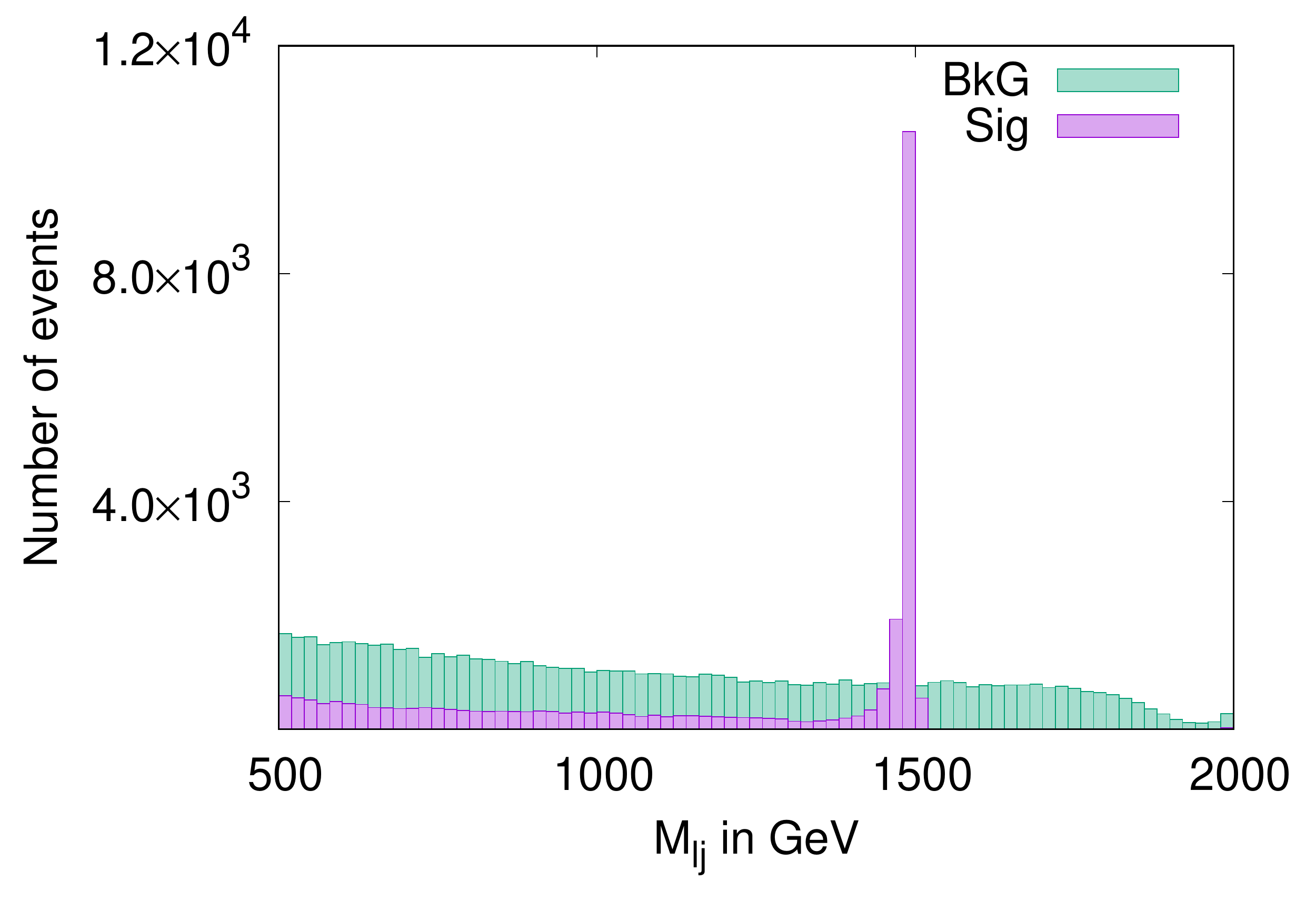}
	
	\includegraphics[width=0.3\textwidth]{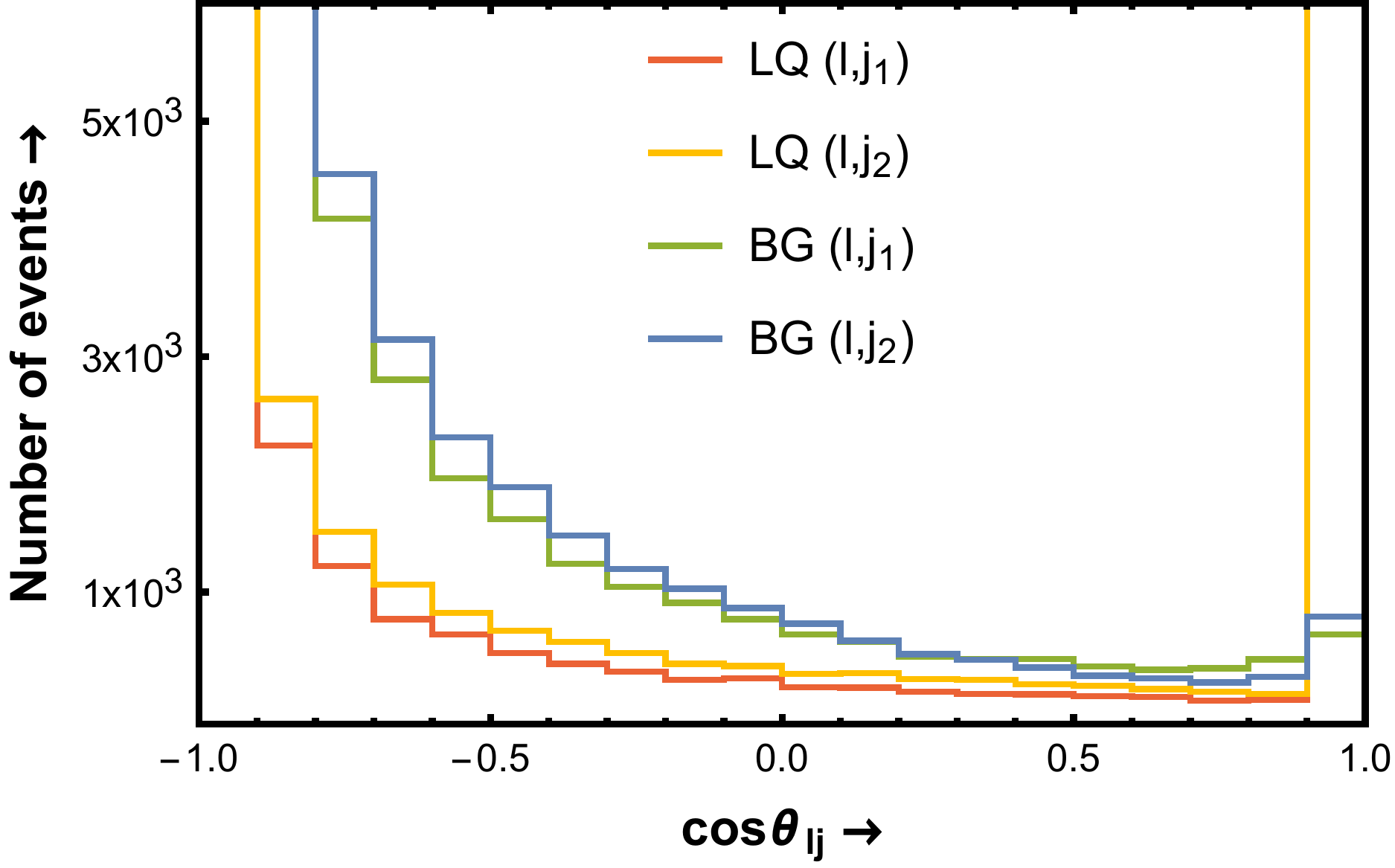}
	\includegraphics[width=0.3\textwidth]{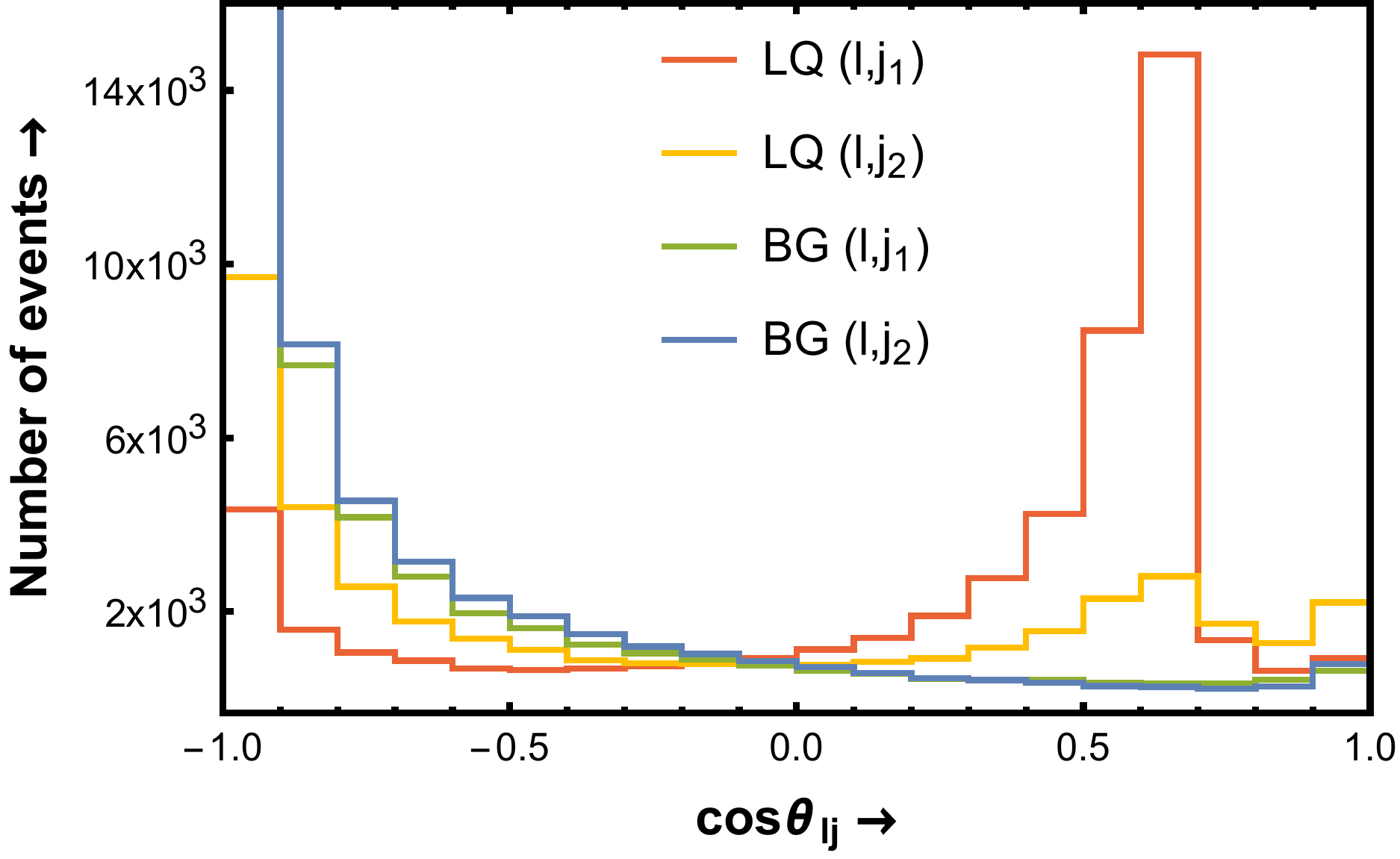}
	\includegraphics[width=0.3\textwidth]{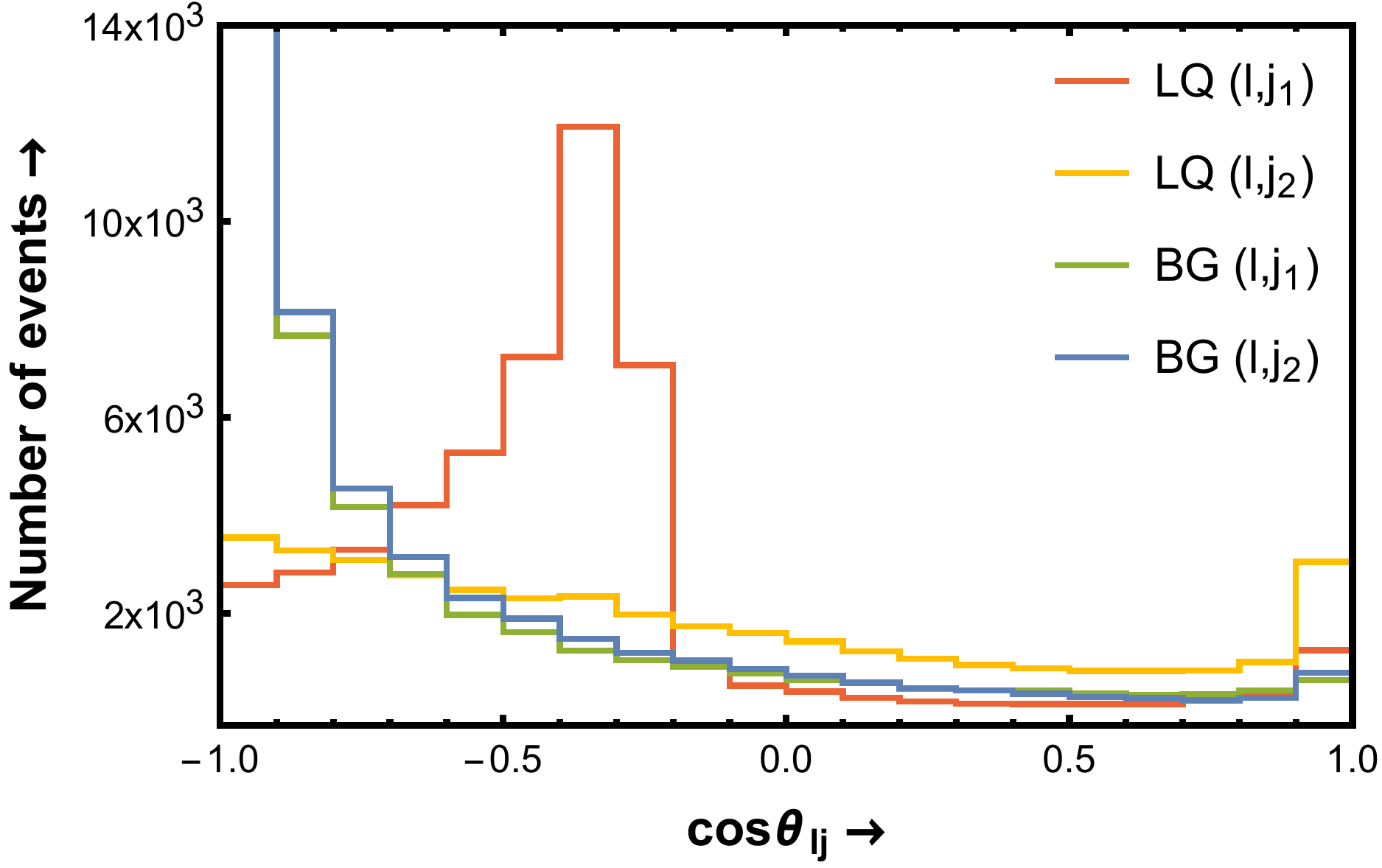}
	\caption{Signal-background simulation for $S_1$ at $e\gamma$ collider with 3 TeV energy of collision, 100 fb$^{-1}$ of integrated luminosity and $10^5$ number of events for three different letpoquark masses: 70 GeV (first column), 650 GeV (second column) and 1500 GeV (third column) \cite{Bandyopadhyay:2020klr}. The first row signifies the invariant mass distributions of jet-lepton pair. In the second row, we present angular distributions of the lepton with the $p_T$ ordered leading and subleading jets, $j_1$ and $j_2$ respectively for the signal and SM backgrounds. The red and yellow lines represent the signal events whereas the green and blue indicate the background events.} \label{Fig:Dis3TeV}
\end{figure}

We finally discuss our analysis in context of the associated production of a leptoquark with a quark in $e\gamma$ collision \cite{Bandyopadhyay:2020klr}. In this section, we observe the scattering angle between the photon and the quark (or anti-quark), produced in association with the leptoquark, or equivalently between the electron and the leptoquark, reconstructed from its decay to charged leptons and quark, in the rest frame of interaction. As discussed in Sec. \ref{Sec:RAZ}, the leptoquarks with charge $\lvert Q_\phi \rvert < 1$ only exhibit zero in the angular distribution in this context. Hence, the leptoquarks with electric charge $-1/3$ (i.e. charge conjugates of $S_1, S_3^{1/3}, V_{2\mu}^{1/3}$ and $\widetilde{V}_{2\mu}^{1/3}$) and $-2/3$ (i.e. charge conjugates of $R_2^{2/3}, \widetilde{R}_2^{2/3},U_{1\mu}$ and $U_{3\mu}^{2/3}$) will show the zero while the others will not. It is worth mentioning that the leptoquarks $S_3^{-2/3}, \widetilde{R}_2^{-1/3}, \widetilde{V}_{2\mu}^{-2/3}$ and $U_{3\mu}^{-1/3}$ can never be produced at any $ep$ or $e\gamma$ collider since they do not interact with the charged leptons. Thus, except these four leptoquarks, all the others show zeros in their angular distributions at either of two  ($ep$ or $e\gamma$) colliders, but none exhibits zero at both of them. Here, we present the results involving the associated production of the scalar leptoquarks, namely, $S_1^{1/3}$, $\widetilde{R}_2^{2/3}$ and of the vector leptoquarks, $U_{1\mu}^{2/3}$ and $\widetilde{V}_{2\mu}^{1/3}$ only.

We perform our analysis with different leptoquark masses, 70 GeV (BP1), 650 GeV (BP2) and 1.5 TeV (BP3) with Yukawa couplings within the allowed parameter space, at different energies of interaction, 200 GeV, 2.0 TeV and 3.0 TeV. Since the electrons and photons are fundamental particles and they collide with identical energies, the interactions registered are sans boost, and are already at the rest frame. As illustrated by Eqn.~\ref{Eq:RAZ2} and Figure \ref{Fig:RAZEGamCos}, the position of the zero in scattering amplitude varies with the leptoquark mass and the interaction energy.

For the simulation, we consider, unlike in the preceding sections, leptoquark decay to all possible channels involving charged leptons and quarks, and hence have considered all irreducible SM backgrounds leading to at least $1\ell + 2 j$ topology to evaluate the signal significances. We therefore demand every events with $1\ell^- + 2j$, evaluate the invariant mass $M_{\ell j}$ for all possible jet-lepton pairs which peaks at $M_\phi$, and demand exactly one event with $\lvert M_{\ell j} - M_\phi \rvert \leq 10$ GeV. In order to increase the signal significance over the background, we then impose a further cut on the angle between the lepton and the jet originating from the leptoquark decay, depending on the estimated boost of the leptoquark, which in turn depends on the leptoquark mass and the interaction energy. In Figure~\ref{Fig:Dis3TeV}, we present the invariant mass distributions and angular distributions of the lepton with respect to the leading and subleading $p_T$ ordered jets for both signal ($S_1$) and background in order to justify the imposition of the cuts.

\begin{figure}[!htb]
	\centering
	\mbox {
		\subfigure[$S_1^{-1/3}$]{\includegraphics[width=0.48\textwidth,height=0.35\textheight]{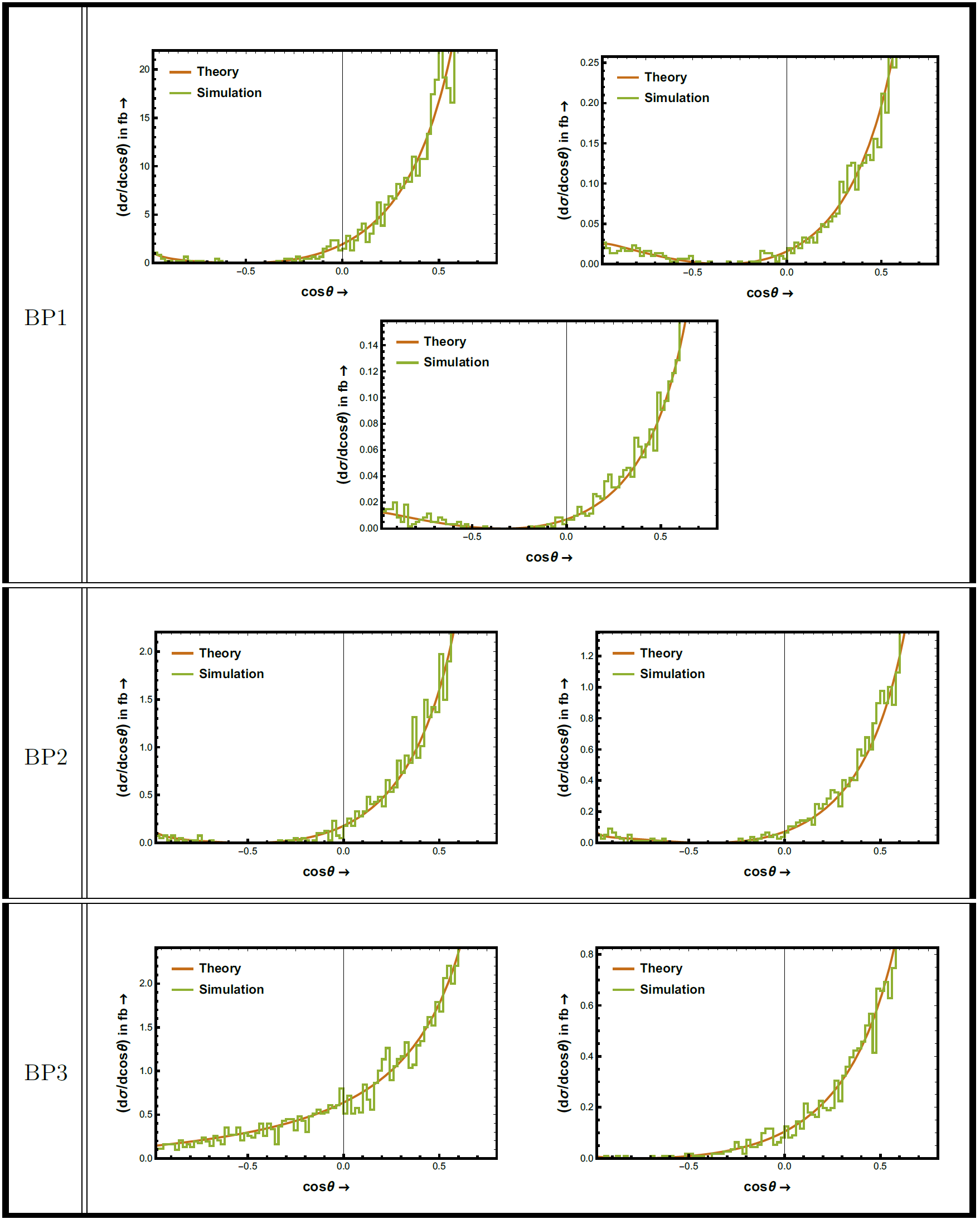}}
		\subfigure[$\widetilde{R}_2^{-2/3}$]{\includegraphics[width=0.48\textwidth,height=0.35\textheight]{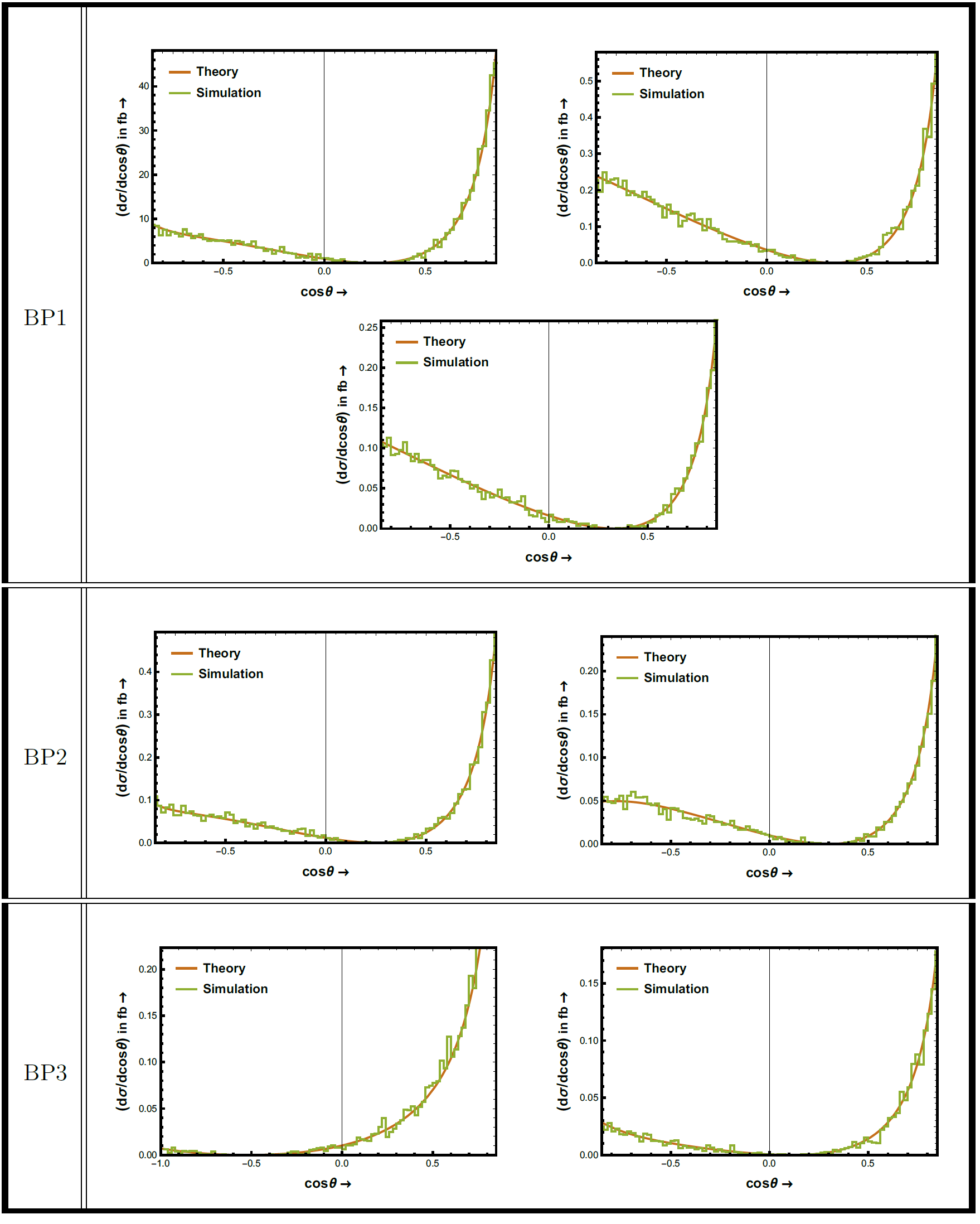}} }
	
	\mbox {
		\subfigure[$U_{1\mu}^{-2/3}$]{\includegraphics[width=0.48\textwidth,height=0.35\textheight]{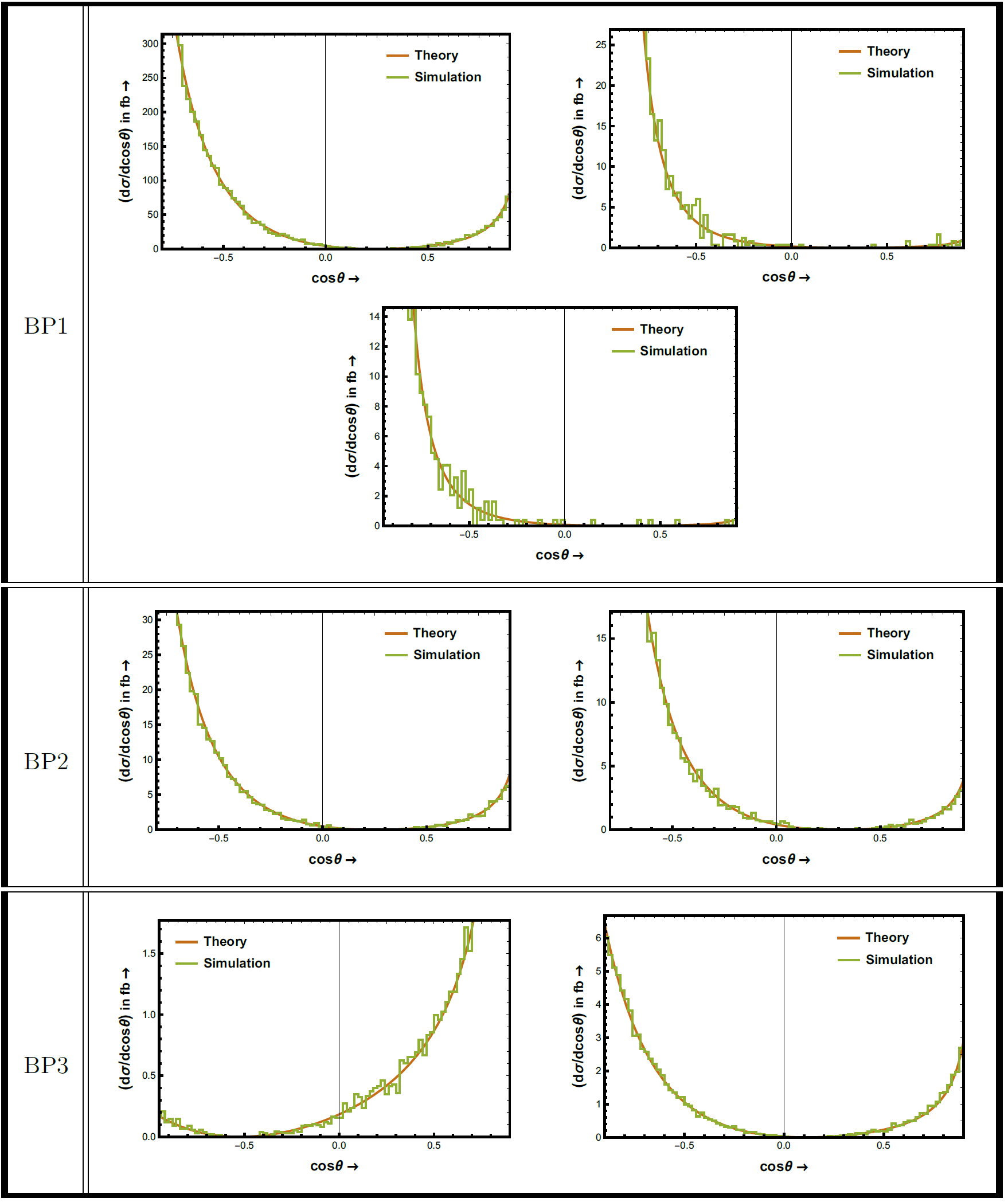}}
		\subfigure[$\widetilde{V}_{2\mu}^{-1/3}$]{\includegraphics[width=0.48\textwidth,height=0.35\textheight]{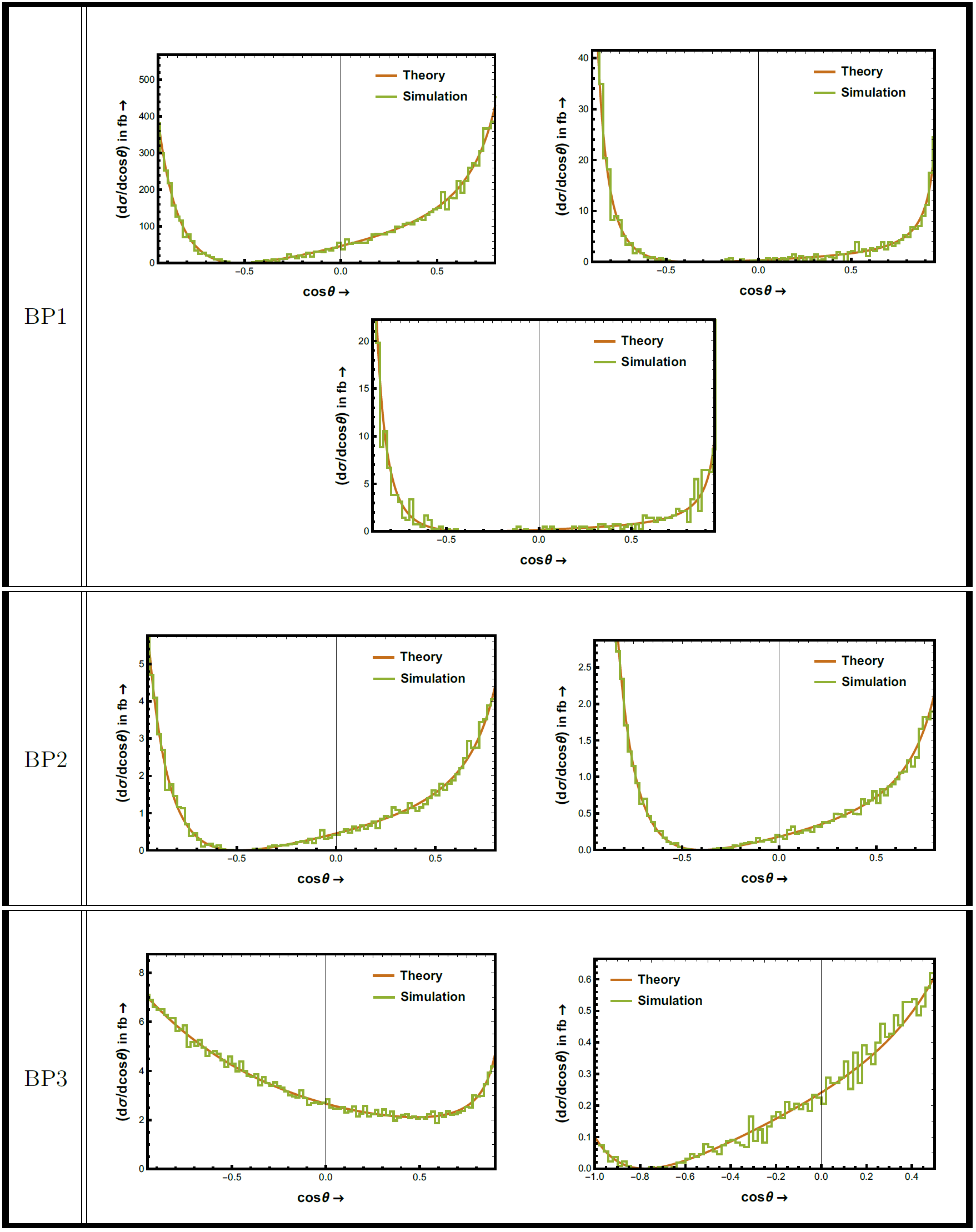}} }
	\caption{Angular distribution for the productions of $S_1^{-1/3},~\widetilde{R}_2^{-2/3},~U_{1\mu}^{-2/3}$ and $\widetilde{V}_{2\mu}^{-1/3}$ at various interaction energies for different BPs \cite{Bandyopadhyay:2020klr}. The brown (smooth) curves indicate the theoretical expectations whereas the green (jagged) lines signify the PYTHIA simulated data with monochromatic photon source.} \label{Fig:RAZCompEGam}
\end{figure}

In Figure~\ref{Fig:RAZCompEGam}, we present a comparison between theoretical prediction and PYTHIA estimate of the angular distribution of the scattered quark, produced in association with the leptoquarks of different mass and $SU(2)_L$ representation, with respect to the incoming photon at different collision energies. We have shown the results for $S_1^{-1/3}$, $R_2^{-2/3}$, $U_{1\mu}^{-2/3}$ and $\widetilde{V}_{2\mu}^{-1/3}$ respectively. Each figure has three subfigures, the first shows the angular distribution of BP1 at all the collision energies while the rest two shows that of BP2 and BP3 respectively at 2.0 and 3.0 TeV collisions, since they cannot be produced at 200 GeV collision. One can observe the agreement between the theoretical prediction and the simulation in each of the plots.

\begin{figure}[!htb]
\centering
\mbox {
\subfigure[]{\includegraphics[width=0.3\textwidth]{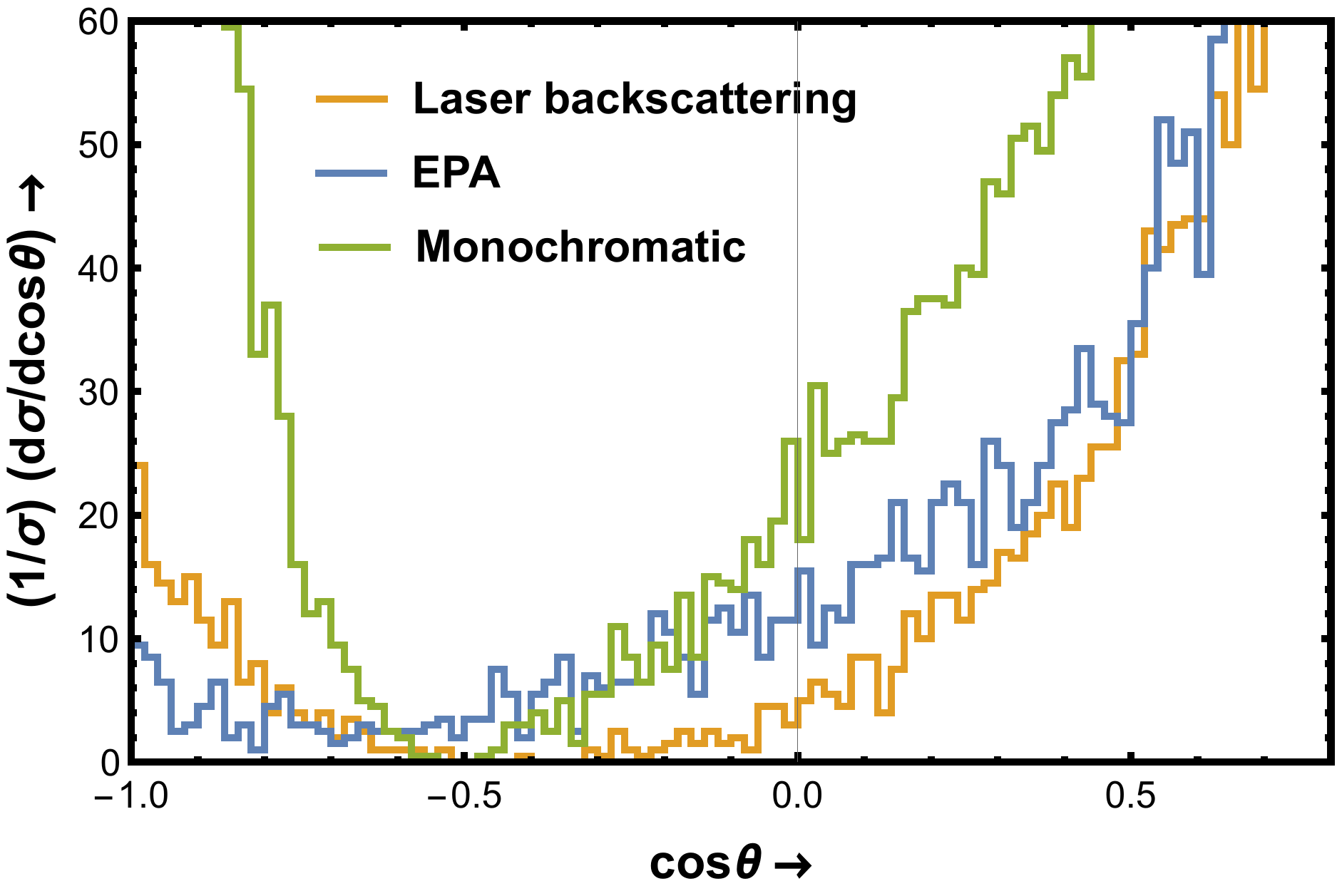}}
\hspace*{2.0cm}
\subfigure[]{\includegraphics[width=0.3\textwidth]{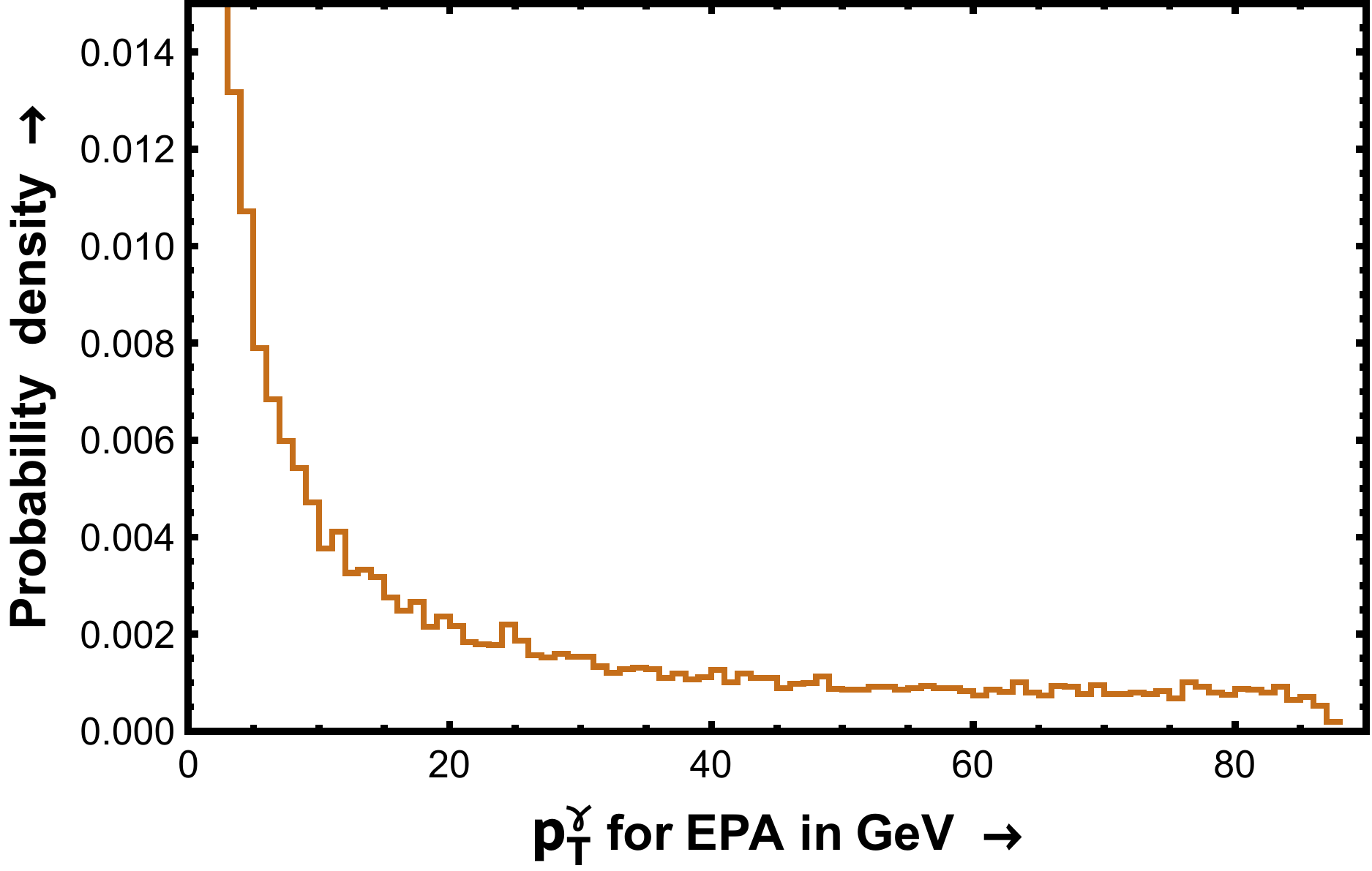}} }
\caption{The comparison between different photon sources, namely, laser backscattering, EPA and monochromatic photons (in orange, blue and green respectively) \cite{Bandyopadhyay:2020klr} on the normalised angular distribution of the scattered 70 GeV leptoquark $\widetilde{V}_{2\mu}^{-1/3}$ with respect to initial electron in terms of weighted differential distribution at $\sqrt{s} = 0.2$ TeV as shown in (a). The distribution for transverse momentum of photon from 100 GeV positron under EPA scheme \cite{Bandyopadhyay:2020klr} is shown in (b).} \label{Fig:DiffGamSrc}
\end{figure}

Nonetheless, our analysis so far have been performed with monochromatic photon source only. But the technology till date is incapable of producing such high energetic monochromatic photons with high intensities. Rather laser backscattering and equivalent photon approximation (EPA) are used as a sources of photons which have some considerable width in wavelength. Now, we briefly present a comparative analysis for different photon sources on the manifestation of the zero in angular distribution of the scattered leptoquark in the left panel of Figure~\ref{Fig:DiffGamSrc}. The weighted differential distributions ($\sigma \frac{d\sigma}{d\cos\theta}$) for the associated production of 70 GeV $\widetilde{V}_{2\mu}^{-1/3}$ at $\sqrt{s} = 0.2$ TeV from laser backscattering, EPA and monochromatic photons are shown in the subfigure (a). As anticipated, the three distributions do not coincide. The laser backscattering preserves the zero of the angular distribution, slightly deviated from the monochromatic case, owing to variation in $\sqrt{s}$ for each collision caused by the distribution of photon energy. For the EPA scheme, the zero gets smeared off following the non-zero distribution of photonic transverse momentum ($p_T^\gamma$) shown in right panel of Figure~\ref{Fig:DiffGamSrc}.

\section{Conclusion and Outlook} \label{Sec:Concl}
As the extensive study suggests the angular distribution of scattered states is an effective observable to discern the spin, gauge representation as well as the electromagnetic charge of the BSM states at present and future colliders. First we consider pair production of leptoquarks at $pp$ collider. From total cross-section and angular distribution one can identify the spin of leptoquark. Vector leptoquarks have much higher production  cross-section and concave-shaped angular distribution plot in contrast to the scalar ones which have very low cross-section and convex-shaped angular distribution graph. In order to differentiate different components of same $SU(2)_L$ multiplet, one should look for determining the charge of the jet produced from the decay of the leptoquark. Next, we look at RAZ for the associated production of a leptoquark with a photon at $ep$ collider. It turns out that the leptoquarks with $|Q_\phi|>1$ would show zeros in the angular distribution relative to the angle between photon and electron. At the end, we study the zeros of single photon tree-level amplitude for associated production of a leptoquark with a quark (or anti-quark) in $e\gamma$ collision. Except $S_3^{-2/3}, \widetilde{R}_2^{-1/3}, \widetilde{V}_{2\mu}^{-2/3}$ and $U_{3\mu}^{-1/3}$, which can never be produced at any $ep$ or $e\gamma$ collider, all other leptoquarks with $|Q_\phi|<1$ would show the zero in this case and thus it becomes complementary to $ep$ collider. It is also important to mention that unlike $ep$ collider, the position of zeros in this case depends not only on the charge of leptoquark but also on the ratio of leptoquark mass to the energy of collision in CM frame. We have also looked at the effects of non-monochromatic photons on the angular distribution. These studies can be used in present and future colliders to probe or rule out different leptoquark scenarios. Similar studies can also also be performed on other BSM particles too.

\section*{Acknowledgements}
We thank SERB CORE Grant CRG/2018/004971 and MATRICS Grant MTR/2020/000668 for the financial support. We also thank Prof. Torbj\"orn Sj\"ostrand for clarification about transverse boost of the initial states in PYTHIA8.

%%%%%%%%%%%%%%%%%%%%%%%%%%%%%%%%%%%%%%%%%%%%%%%%%%%%%%%%%%%%%%%%%%%%
%\newpage
\bibliographystyle{unsrt}

\end{document}